\documentclass[a4paper,fleqn,usenatbib]{mnras}

\usepackage{newtxtext,newtxmath}
\usepackage[T1]{fontenc}
\usepackage{ae,aecompl}

\usepackage{graphicx}	% Including figure files
\usepackage{amsmath}	% Advanced maths commands
\usepackage{amssymb}	% Extra maths symbols%
\usepackage{color}
\usepackage{fancybox}
\usepackage{subfigure}
\usepackage{rotating}
\usepackage{import}
\usepackage{pdfpages}
\usepackage{xcolor}
\usepackage{hyperref}
\usepackage{verbatim} 
\usepackage{CJK}
\usepackage{etoolbox}
\makeatletter
\patchcmd\@combinedblfloats{\box\@outputbox}{\unvbox\@outputbox}{}{%
   \errmessage{\noexpand\@combinedblfloats could not be patched}%
}%
 \makeatother
\usepackage{adjustbox}

\usepackage[switch,columnwise]{lineno}
\newcommand{\angstrom}{{\rm \AA}}

\newcommand{\kms}{$\rm km~s^{-1}$}

\newcommand{\hb}{H{$\beta$}}
\newcommand{\ha}{H{$\alpha$}}

\begin{document}

\title[SEDs of candidate periodic quasars]{Spectral Energy Distributions of Candidate Periodically-Variable Quasars: Testing the Binary Black Hole Hypothesis}

% The list of authors, and the short list which is used in the headers.
% If you need two or more lines of authors, add an extra line using \newauthor

\author[H. Guo et al.]{
Hengxiao Guo,$^{1,2}$\thanks{E-mail: hengxiao@illinois.edu (HG), xinliuxl@illinois.edu (XL)}
Xin Liu,$^{1,2}$
Zafar Tayyaba,$^{3}$ 
Wei-Ting Liao$^{1,2}$\\
% List of institutions
$^{1}$Department of Astronomy, University of Illinois at Urbana-Champaign, Urbana, IL 61801, USA\\
$^{2}$National Centre for Supercomputing Applications, University of Illinois at Urbana-Champaign, 605 East Springfield Avenue, Champaign, IL 61820, USA \\
$^{3}$Australian Astronomical Observatory, PO Box 915, North Ryde, NSW 1670, Australia \\
%$^{4}$ Alfred P. Sloan Research Fellow \\
}

% These dates will be filled out by the publisher
\date{Accepted XXX. Received XXX; in original from XXX}

% Enter the current year, for the copyright statements etc.
\pubyear{2019}

% Don't change these lines

\label{firstpage}
\pagerange{\pageref{firstpage}--\pageref{lastpage}}
\maketitle

% Abstract of the paper
\begin{abstract}

Periodic quasars are candidates for binary supermassive black holes (BSBHs) efficiently emitting low frequency gravitational waves. Recently, $\sim$150 candidates were identified from optical synoptic surveys. However, they may be false positives caused by stochastic quasar variability given the few cycles covered (typically 1.5). To independently test the binary hypothesis, we search for evidence of truncated or gapped circumbinary accretion discs (CBDs) in their spectral energy distributions (SEDs). Our work is motivated by CBD simulations that predict flux deficits as cutoffs from central cavities opened by secondaries or notches from minidiscs around both BHs. We find that candidate periodic quasars show SEDs similar to those of control quasars matched in redshift and luminosity. While seven of 138 candidates show a blue cutoff in the IR-optical-UV SED, six of which may represent CBDs with central cavities, the red SED fraction is similar to that in control quasars, suggesting no correlation between periodicity and SED anomaly. Alternatively, dust reddening may cause red SEDs. The fraction of extremely radio-loud quasars, e.g., blazars (with $R>100$), is tentatively higher than that in control quasars (at 2.5$\sigma$). Our results suggest that, assuming most periodic candidates are robust, IR-optical-UV SEDs of CBDs are similar to those of accretion discs of single BHs, if the periodicity is driven by BSBHs; the higher blazar fraction may signal precessing radio jets. Alternatively, most current candidate periodic quasars identified from few-cycle light curves may be false positives. Their tentatively higher blazar fraction and lower Eddington ratios may both be caused by selection biases.  
\end{abstract}

% Select between one and six entries from the list of approved keywords.
% Don't make up new ones.

%%%%%%%%%%%%%%%%%%%%%%%%%%%%%%%%%%%%%%%%%%%%%%%%%%

%%%%%%%%%%%%%%%%% BODY OF PAPER %%%%%%%%%%%%%%%%%%
%%%%%%%%%%%%%%%%%%%%%%%%%%%%%%
\section{Introduction}

The observed growths of structures suggest that mergers of galaxies, and by extension, their central supermassive black holes \citep{kormendy95,KormendyHo2013} (SMBHs), should be common throughout most of cosmic history \citep[e.g.,][]{begelman80,kauffmann00,milosavljevic01,haehnelt02, yu02,volonteri03,hopkins08,Blecha2013a,Steinborn2015}. Low frequency gravitational waves (GWs) are expected from the final coalescence of merging SMBHs. As a major target in the emerging new field of gravitational astronomy, binary supermassive black holes (BSBHs) provide a ``standard siren'' for cosmology and a direct test-bed for strong-field general relativity \citep[e.g.,][]{hughes09,Centrella2010,colpi09,Dotti2012,Tamanini2016}. Unlike stellar mass binary black holes (which are advanced LIGO's primary targets, e.g., \citealt{LIGO2016}) whose detection is largely limited to the local Universe, merging BSBHs would be detectable almost close to the edge of the observable Universe \citep[e.g., z $>$ 7,][]{Klein2016}. The more massive, low-redshift population (i.e., in the relatively nearby Universe) is being hunted by pulsar timing arrays \citep[e.g.,][]{Zhu2014,Shannon2015,Babak2016,Simon2016,Dvorkin2017,Kelley2017a,Mingarelli2017,Wang2017c,Aggarwal2018,Holgado2018,Sesana2018}, whereas the less massive, high-redshift population (i.e., in the earlier Universe) will be targeted by space-borne experiments in future \citep[e.g.,][]{Babak2011,Amaro2012}.

While the formation of BSBHs seems inevitable, direct evidence has been elusive. No confirmed case is known in the GW-dominated regime, where a binary is so close that the orbital decay is driven by emitting GWs. A critical issue is that the orbital decay of a BSBH may significantly slow down or even stall at $\sim$parsec scales, i.e., the so called ``final-parsec'' problem \citep{begelman80,milosavljevic01,yu02,DEGN}. There may be a bottleneck when a binary runs out of stars to interact with, yet the gravitational wave emission is still too weak to merge the binary within the age of the universe. This bottleneck represents the largest uncertainty on the abundance of BSBH mergers as low-frequency GW sources. In theory, the bottleneck may be overcome in gaseous environments \citep[e.g.,][]{gould00,Cuadra2009,Chapon2013,delValle2015}, in triaxial or axisymmetric galaxies \citep[e.g.,][]{Khan2016,Kelley2017}, and/or by interacting with a third BH in hierarchical mergers \citep[e.g.,][]{blaes02,Kulkarni2012,Bonetti2018}. 

Observational searches for BSBHs are important for testing different orbital evolutionary theories and their efficiency in solving the final-pc problem. However, typical physical separations of BSBHs that are gravitationally bound to each other ($\lesssim$ a few pc) are too small for direct imaging. Even VLBI cannot resolve BSBHs except for in the local universe \citep{burke11}. CSO 0402+379 (discovered by VLBI as a double flat-spectrum radio source separated by 7 pc) remains the only robust case known \citep{rodriguez06,Bansal2017}. While great strides have been made in identifying dual active galactic nuclei -- progenitors of BSBHs at $\gtrsim$kpc scales \citep[e.g.,][]{komossa03,ballo04,hudson06,Liu2013,Comerford2015,Fu2015a,Muller-Sanchez2015,Koss2016,Ellison2017,Liu2018b,Hou2019}, there is no consensus case of BSBHs at millipc scales, i.e., in the GW regime \citep[e.g.,][]{Bogdanovic2015,Komossa2015a}. Indirect searches are needed to idenfity BSBHs beyond the local universe.

Periodic quasar light curves have long been suggested as indirect evidence for candidate millipc BSBHs. The optical flux periodicity may be caused by accretion rate changes due to the intrinsic binary orbital tidal torque modulation \citep[e.g.,][]{MacFadyen2008,Shi2012,Roedig2012,DOrazio2013,Farris2014,Tang2018}, and/or the apparent Doppler boost modulation from the highly relativistic motion of gas in the mini accretion disc around the secondary BH \citep[e.g.,][]{DOrazio2015a,Charisi2018}. 
While $\gtrsim$100 quasars with candidate periodicity have been proposed as evidence for BSBHs \citep[e.g.,][]{valtonen08,Graham2015,Graham2015a,Liutt2015,Liutt2016,Bon2016,Charisi2016,Zheng2016,Li2019}, even the strongest candidate periodicity has been shown to be subject to false positives due to stochastic quasar variability given the uneven sampling, limited time baseline, and/or relatively low sensitivity. For example, the blazar OJ 287 has been suggested to host a BSBH based on the evidence for a 12-year periodicity in the optical and radio light curves \citep[e.g.,][]{sillanpaa96,valtaoja00,Valtonen2016}, where the double-peaked flares have been interpreted as the result of a secondary BH punching through the accretion disc of the primary \citep[e.g.,][]{Takalo1994,valtonen08} or accretion disc precession driven by the gravitational torque of a companion BH \citep[e.g.,][]{katz97}. However, \citet{Goyal2018} has shown that out of the 117-year total duration of the available optical light curves, the observations before 1970 were highly irregularly sampled, whereas the better-sampled 1970--2017 light curve covers only $\sim$3 of the claimed cycles and is too short to detect any significant periodicity over the coloured-noise (i.e., stochastic component) of the power spectrum. Another example is the blazar PG1302$-$102, originally proposed as a BSBH candidate based on evidence for a 5-year periodicity \citep{Graham2015a} and interpreted as due to relativistic Doppler boost \citep{DOrazio2015}, which has been suggested to be a false positive from random quasar variability \citep[e.g.,][but see \citealt{Kovavcevic2019}]{Vaughan2016,Liutt2018}. Furthermore, even if the suggested periodicity were true, the physical mechanism driving the periodicity is still uncertain \citep[e.g.,][]{Graham2015,Charisi2018}. In addition to BSBHs, alternative scenarios may be responsible for driving optical periodicity, including warped accretion discs \citep[e.g.,][]{Tremaine2014}, radio jet procession \citep[e.g.,][]{Kudryavtseva2011,Caproni2017,Sobacchi2017}, quasi-periodic oscillations (QPOs) from e.g., Lens-Thirring procession \citep[e.g.,][]{Stella1998,Ingram2011}, and resonant accretion of magnetic field lines \citep[i.e., ``magnetic breathing'' of the accretion disc; e.g.,][]{Villforth2010}. Complementary tests are needed to verify any candidate periodicity and to sort out alternative scenarios for its physical origin.

In this work, we search for evidence of a truncated or gapped circumbinary accretion disc in a sample of candidate periodic quasars compiled from the literature by studying their spectral energy distributions (SEDs). Given the typical $\sim$yearly cycles and the total black hole masses (${\sim}10^8$--$10^9M_{\odot}$) of the known candidate periodic quasars, the claimed BSBHs are generally expected at pre-decoupling \citep[e.g.,][]{Kocsis2011a,Tanaka2012,Sesana2012}, i.e., when the gravitational wave inspiral timescale is still longer than the viscous timescale, where circumbinary accretion discs should be common. The current work is motivated by circumbinary accretion disc models that predict abnormalities such as a cutoff or notch in the IR-optical-UV SED, depending on the mode of circumbinary accretion and the evolutionary state of the system \citep[e.g.,][]{milosavljevic05,Gultekin2012,Kocsis2012a,Tanaka2013,Tanaka2013a,Gold2014a,Roedig2012,Roedig2014,Farris2015,Farris2015a,Krolik2019}. For BSBHs with near-equal mass ratios (e.g., $q{>}0.1$), the secondary BH may open a cavity in the inner region of the circumbinary accretion disc resulting in a cutoff in the SED, or the two BHs may keep accreting gas from the circumbinary disc and maintaining their own minidiscs producing ``notches'' in the SED. By searching for SED abnormalities, our current work serves as a complementary test of the BSBH hypothesis for candidate periodic quasars.

The paper is organized as follows. \S\ref{sec:pred} briefly reviews theoretical predictions of BSBH circumbinary accretion discs, focusing on abnormalities that may be observable in the optical/UV SEDs. \S\ref{sec:sample} then describes the sample of candidate periodic quasars compiled from the literature and the SED data from available archival observations, as well control samples of ordinary quasars to put the results of candidate periodic quasars into context. \S\ref{sec:sed} presents the SED properties of candidate periodic quasars in comparison to control quasars and the identification of a sample of seven candidate periodic quasars that show apparent SED abnormalities. \S\ref{sec:dis} discusses the possible physical origins of SED abnormalities of the sample of seven candidate periodic quasars, highlighting in particular the internal reddening due to dust in the quasar host galaxies. Finally, we summarize our main results and conclude in \S\ref{sec:con}. Throughout this paper, we assume a $\Lambda$CDM cosmology with $H_{0}$ = 70 $\rm km\,s^{-1}Mpc^{-1}$, $\Omega_{m}$ = 0.3, and $\Omega_{\Lambda}$ = 0.7.

%%%%%%%%%%%%%%%%%%%%%%%%%%%%%%%%%%%%%%%%%%%%%%%%%%%%%%%%%%%%%%%%%%%%%%%%%%%%%%%%%%%%%%%%%%%%%%%%
\section{Theoretical Predictions of BSBH Circumbinary Accretion disc SEDs}\label{sec:pred}

\begin{figure}
 \centering
  \hspace*{-0.8cm}
  \includegraphics[width=100mm]{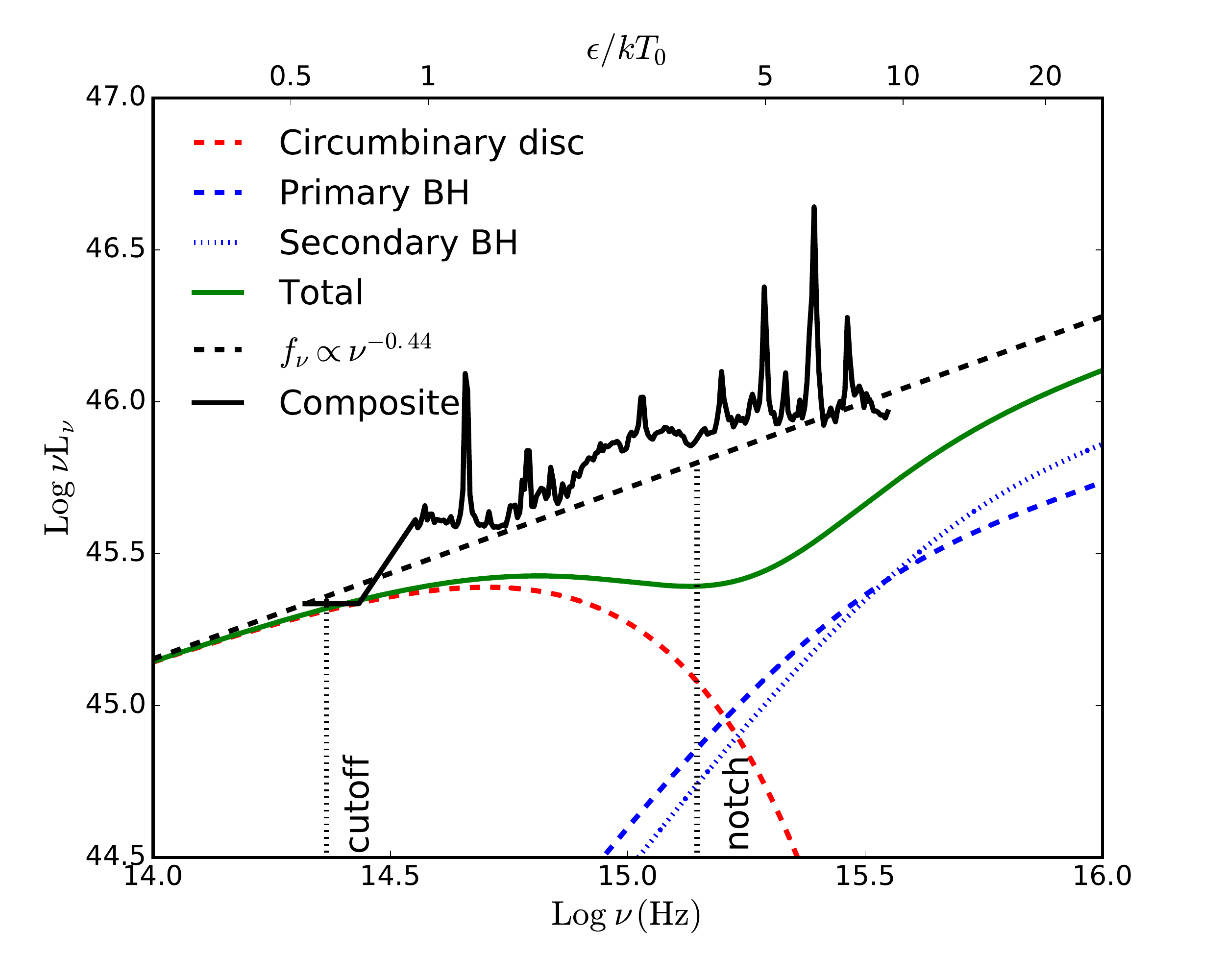}
    \caption{Illustration of theoretical SEDs of BSBH circumbinary accretion discs. 
    The red dashed curve shows redder (colder) thermal emission from a circumbinary disc, whereas the blue dashed and dotted represent bluer (hotter) emission from minidiscs around the primary and secondary BHs, if any. The green solid shows the total thermal emission, assumed to be the sum of the radiation from the circumbinary disc and that from the two minidiscs. 
    We assume that the total BH mass is $10^{8}M_{\odot}$ with a mass ratio of $q = 0.1$. The total high-frequency emission from the gas surrounding the primary (blue dashed) and secondary (blue dotted) BHs are 45\% and 55\%, respectively.
    The dotted lines mark the expected locations of the SED cutoff and north. The solid black curve shows the SDSS quasar composite spectrum \citep{VandenBerk2001}, whose continuum is fit with a power-law over the spectral range of 1300--5000 {\AA} (black dashed). See \S \ref{sec:pred} for details.
}
  \label{fig:theory}
\end{figure}

Figure \ref{fig:theory} illustrates theoretical SEDs of BSBH circumbinary accretion discs in the IR-optical-UV. Models of BSBH circumbinary accretion discs predict two characteristic morphologies that may indicate the presence of BSBHs through abnormalities in their IR-optical-UV SEDs \citep[e.g.,][]{Roedig2014,Foord2017,Tang2018}. One is a central cavity, where the inner region of the circumbinary disc is almost emptied by the secondary BH. For BSBHs with near-equal mass ratios (e.g., $q{>}0.1$), the emission would be truncated blueward of the wavelength that corresponds to the temperature of the innermost disc edge \citep[e.g.,][]{Gultekin2012,DOrazio2013}, producing a sharp exponential cutoff in the IR-optical-UV SED as illustrated in Figure \ref{fig:theory} (the blue dashed curve). The other is minidiscs, where there is substantial accretion onto one or both BHs, each with their own shock-heated thin disc \citep[e.g.,][]{Yan2015,Ryan2017,Tang2018}. The minidiscs emit high energy radiation analogous to a single BH with a geometrically thin and optically thick disc (the red dashed and dotted curves in Figure \ref{fig:theory}). In this scenario, the emergence of a gap between the tidal radii of the minidiscs and truncation radius of the circumbinary disc will lead to a notch in the SED of the total emission (the green solid curve in Figure \ref{fig:theory}).

The location of the flux deficit primarily depends on the temperature of the inner edge of the circumbinary disc (i.e., the ``cutoff'' temperature) given by
\begin{equation}\label{eq:T}
T_{{\rm cutoff}} \simeq 2.0 \times10^{4} \bigg[\dot{m}\bigg(\frac{\eta }{0.1}\bigg)^{-1} M_{8}^{-1} \bigg(\frac{a}{100 R_{g}}\bigg)^{-3}\bigg]^{\frac{1}{4}} \, {\rm K},
\end{equation}
where $\dot{m}\equiv \dot{M}/\dot{M_{{\rm Edd}}}$ is the accretion rate in Eddington units, $\eta$ is the radiative efficiency, $M_{8} \equiv M_{{\rm BH}}/(10^8M_{\odot})$ is the total binary BH mass in units of $10^{8}M_{\odot}$, $a$ is the binary's semimajor axis, and $R_{g} \equiv GM/c^2$ is the gravitational radius \citep{Roedig2014}. Figure \ref{fig:theory} shows a typical example where we assume $\dot{m}=0.1$, $M_{8}=1$, $\eta=0.1$, and $a=50R_{g}$, resulting in $T_{{\rm cutoff}} \simeq 11,000$ K, which corresponds to 2600 \AA\ based on Wien's law. The deepest portion of the notch happens around $T_{{\rm notch}}\simeq4T_{0}$, where $T_{0}= 2^{3/4}T_{{\rm cutoff}}$ is the characteristic temperature of the accretion disc of a single BH with $r{\sim}a$ that lies between the hottest temperature of the circumbinary disc \citep[truncated at ${\sim}2a$;][]{Farris2014} and the coldest temperature in the minidiscs \citep[extended to ${\sim}a/2$;][]{Paczynski1977}. The entire notch ranges from about $\epsilon \equiv kT\simeq$1 to 15 $kT_{0}$, where $k$ is the Boltzmann constant. The exact width and depth of the notch depend on the binary mass ratio and the relative rate of gas flowing onto the two BHs. The more gas flowing onto the smaller BH, the wider and deeper the notch will be, with its center being relatively more stable. The deepest portion of the notch is at most a factor of ${\sim}3$ fainter than single BH case \citep{Roedig2014}. 

In the simple illustration, we have assumed that the thermal radiation from an accreting BSBH is the sum of the radiation from the circumbinary disc and the radiation from the two minidiscs. We have ignored the contribution of streams in the cavity, which connect the minidiscs and the circumbinary disc, since their contribution to the total light is expected to be less than $\sim$10\% \citep[e.g.,][]{Tang2018}. We have also ignored possible smoothing effect on the notch by gas streams \citep[e.g.,][]{dAscoli2018}.

Figure \ref{fig:theory} also shows the observed IR-optical-UV quasar composite SED constructed based on a sample of $\sim$2000 SDSS quasars \citep{VandenBerk2001}. The power-law index of the observed quasar composite is $\alpha_{\nu,{\rm obs}}=-0.44$ over the spectral range of $\sim$1300--5000 \AA, whereas the theoretical value is $\alpha_{\nu,{\rm th}}=1/3$ predicted based on a multicolour black-body model. The difference may be caused by internal dust reddening in the quasar host galaxies \citep[e.g.,][]{Xie2016}. We have scaled the theoretical power-law index to be consistent with the observed value to mimic the effect of dust reddening.

%In the following sections, we will explore these cutoff and notch signatures in quasar SED to hint any possible SBHB candidates.

%%%%%%%%%%%%%%%%%%%%%%%%%%%%%%%%%%%%%%%%%%%%%%%%%%%%%%%%%%%%%%%%%%%%%%%%%%%%%%%%%%%%%%%%%%%%%%%%
\section{Sample and Data}\label{sec:sample}

\subsection{A Sample of Candidate Periodically-Variable Quasars Compiled from the Literature}

We combine the two largest known samples of systematically selected candidate periodic quasars as described below (\S\ref{subsubsec:crts} \& \S\ref{subsubsec:ptf}). We do not include other individually identified candidates \citep[e.g.,][]{Zheng2016,Li2019} to focus on a more homogeneous sample.

\subsubsection{The CRTS Sample from \citet{Graham2015}}\label{subsubsec:crts}

We include 111 candidate periodic quasars selected by \citet[][hereafter G15]{Graham2015} using data from the Catalina Real-time Transient Survey (CRTS\footnote{http://crts.caltech.edu}). Established in late 2007, the CRTS is a synoptic survey that covers $\sim$33,000 deg$^{2}$ of the sky to discover optical transients \citep{Drake2009,Djorgovski2011}. It uses data automatically collected by the three dedicated 1 m class telescopes of the Catalina Sky Survey near-Earth object project. The CRTS has produced publicly available time series down to a V-band limit of $\sim$20 mag for ${\sim}5\times10^8$ objects with an average of $\sim$250 observations over a 9-year baseline. 

From a parent sample of 243,500 spectroscopically confirmed quasars, G15 identified 111 candidates that show a strong Keplerian periodic signal with at least 1.5 cycles over the 9-year baseline using a joint wavelet and autocorrelation funciton-based approach. Subject to the light curve time baseline and the minimal number of cycles covered, most rest-frame periods of these candidates are around 2$\sim$3 yrs. The blazar PG1302$-$102 \citep{Graham2015a} represents the strongest periodic candidate in the G15 sample.

\subsubsection{The PTF Sample from \citet{Charisi2016}}\label{subsubsec:ptf}

We also consider 33 candidate periodic quasars selected by \citet[][hereafter C16]{Charisi2016} using data from the Palomar Transient Factory \citep[PTF;][]{Rau2009,Law2009}. The PTF was an optical synoptic survey to explore the transient and variable sky. It lasted from 2009/03 to 2012/12. The observations were made at Palomar Observatory by the 1.2 m Samuel Oschin Schmidt telescope with the CHF12K camera, providing a wide field of view of 7.26 deg$^2$. It covered $\sim$3000 deg$^2$ of the sky with a 5$\sigma$ limiting magnitude of $\sim$20.6 in Mould-$R$ and $\sim$21.3 in SDSS-$g$ bands with an average 5 day cadence. 

From a parent sample of 35,383 spectroscopically confirmed quasars, C16 selected 50 candidate periodic quasars with at least 1.5 cycles within the PFT baseline by identifying unusually high peak in the Lomb-Scargle periodograms of the optical light curves, whose statistical significance was assessed by simulating time series that exhibit stochastic damped random walk \citep[][]{Kelly2009,Kozlowski2016} variability. Among the 50 candidates, 33 remain significant with the re-analysis of light curves including data from the intermediate-PTF \citep[iPTF;][]{Cao2016,Masci2017} and CRTS. Of the 33 periodic quasar candidates from the C16 sample, we remove six that have fewer than five bands of archival photometry, resulting in a sample of 27 quasars included in our SED study. The median rest-frame period of these candidates is $\sim$ 1 yr.

The final sample of candidate periodic quasars included in our SED study consists of 138 spectroscopically confirmed quasars (111 from G15 and 27 from C16) in the redshift range of $0<z<3.5$. Figure \ref{fig:para} shows the basic quasar sample properties.

\subsection{Control Sample of Ordinary Quasars}\label{subsec:control}

To put our results into context, we construct a control sample of 1380 ordinary quasars that are matched to have the same redshift and $i$-band absolute magnitude $M_{i,z=2}$ distribution to those of our candidate periodic quasar sample. The control sample was drawn from the SDSS DR14 quasar catalog \citep{Paris2018} and is 10$\times$ the size of the candidate periodic quasar sample. We use KDTree\footnote{https://docs.scipy.org/doc/scipy-0.15.1/reference/generated/\\scipy.spatial.KDTree.query.html}, which looks up the nearest neighbours of any points in the redshift--$M_{i,\,{\rm z=2}}$ space \citep{Maneewongvatana1999}. 90\% sources in the control sample have deviations that are smaller than 0.2 and 0.4 in the redshift and $M_{i, z=2}$ distributions, respectively.

\begin{figure*}
 \centering
 \hspace*{-0.8cm}
  \includegraphics[width=0.8\textwidth]{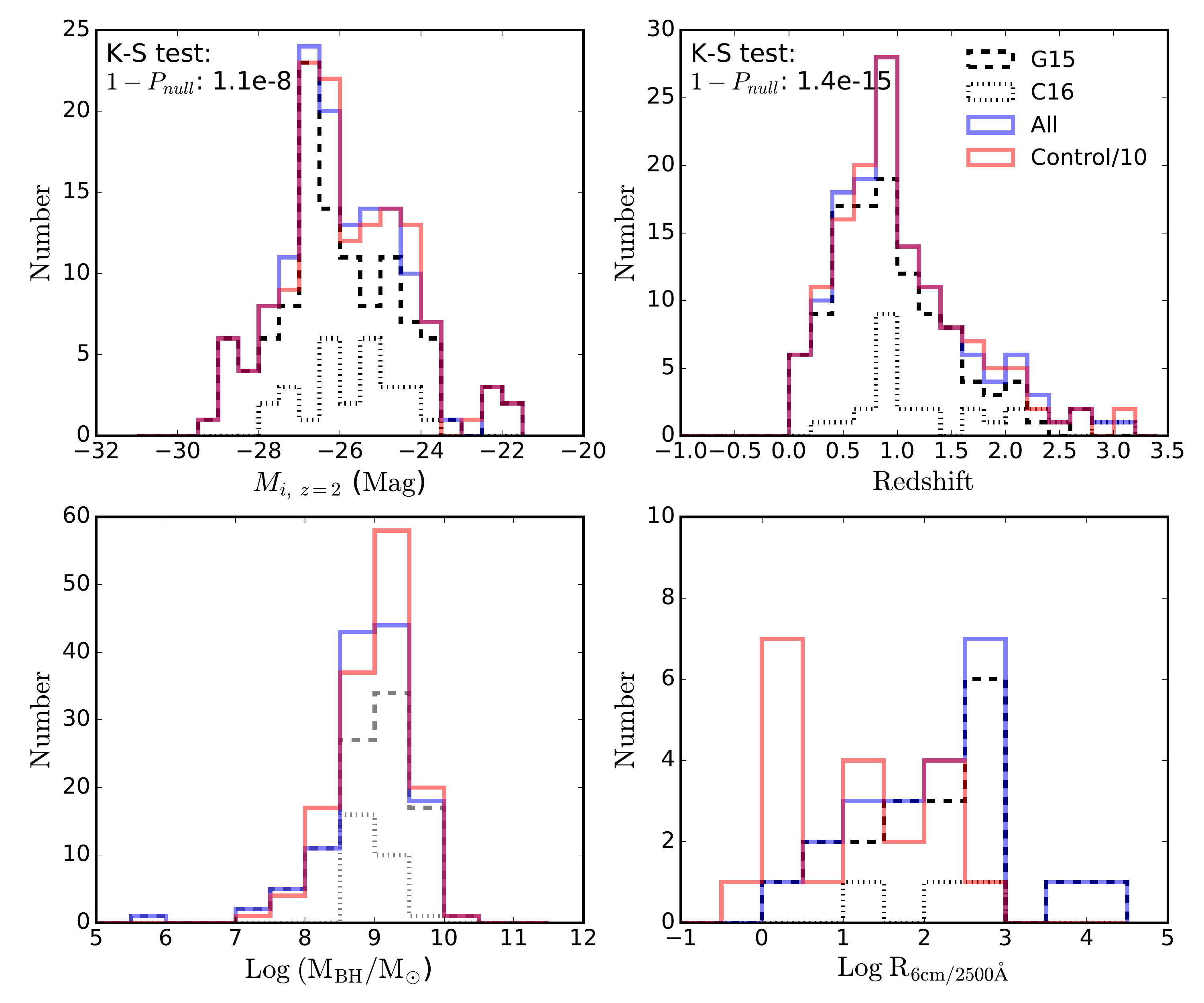}
    \caption{Distributions of $i$-band absolute magnitude, redshift, virial black hole mass, and radio loudness for candidate periodic quasars (138 objects) compared against control quasars (1380 objects, scaled by 1/10). The control quasars were drawn to match the redshift and $i$-band absolute magnitude distributions of candidate periodic quasars. The Kolmogorov-Smirnov (KS) test p-values are labeled on the plot. The fraction of extremely radio-loud quasars, i.e., blazars with $R > 100$, is higher than that in control quasars (at $\sim$2.5$\sigma$).
    }
  \label{fig:para}
\end{figure*}

\begin{figure}
 \centering
 %\hspace*{-0.8cm}
  \includegraphics[width=0.5\textwidth]{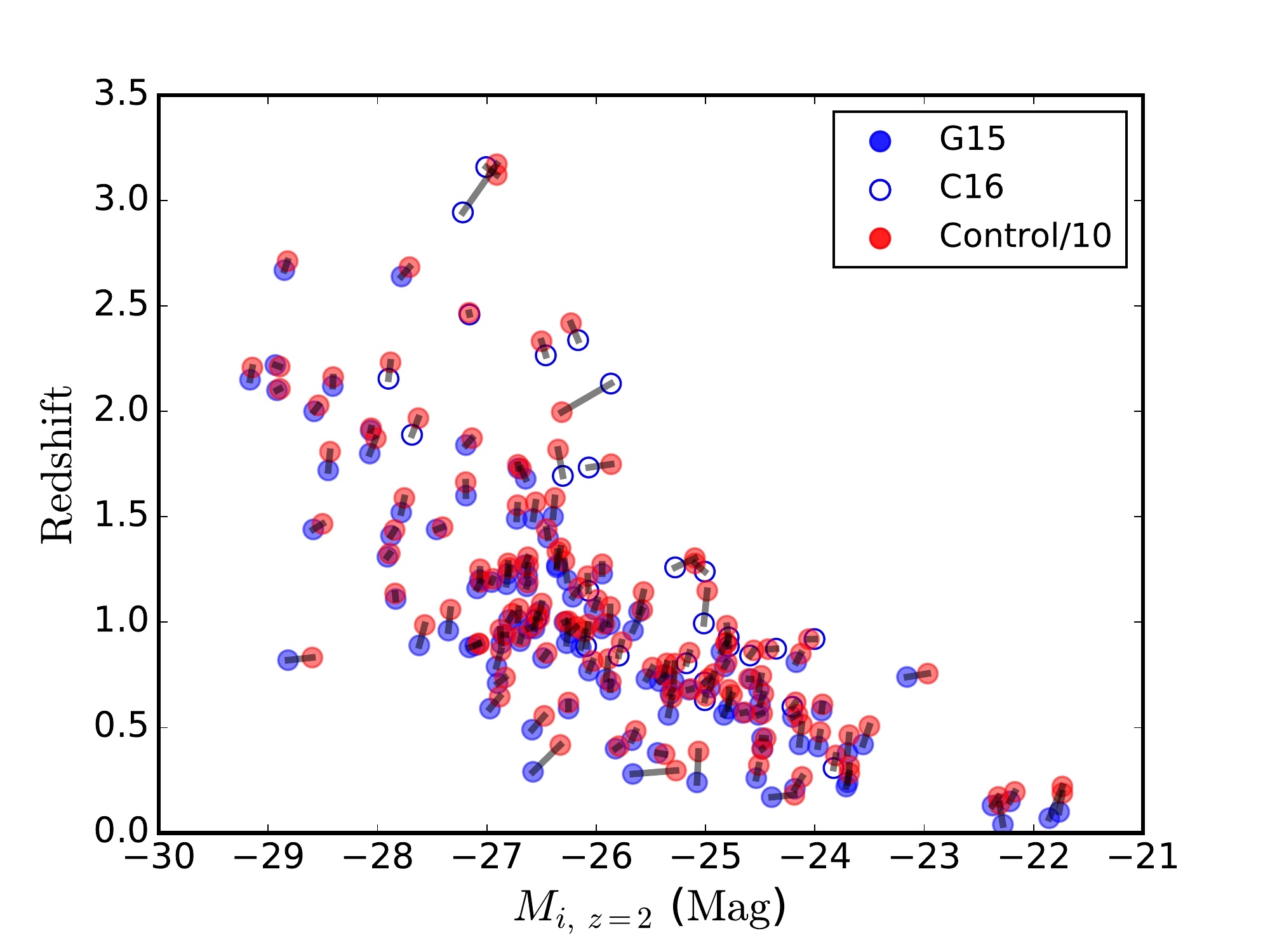}
    \caption{Distributions of $i$-band absolute magnitude and redshift for candidate periodic quasars (138 obejcts) compared against control quasars (a randomly drawn subset of 138 from a total of 1380 objects). The candidates (G15 \& C16) and their corresponding control objects are connected with grey lines.
    }
  \label{fig:M_z}
\end{figure}

Figure \ref{fig:para} shows the redshift and $M_{i,z=2}$ distributions of the candidate periodic quasar sample compared to those of the control sample. As shown in Figure \ref{fig:M_z}, we have double checked that the joint distributions of $M_{i,z=2}$ and redshift are also similar between the periodic and control samples. Also shown in Figure \ref{fig:para} are the distributions of their virial black hole mass estimates $M_{{\rm BH}}$ from \citet{shen11} and \citet{Kozlowski2017} (which will be used to estimate the expected location of the SED notch/cutoff; see discussion below in \S \ref{subsec:sed_location}), and the radio loudness parameter, $R_{6\,{\rm cm}/2500\,\mbox{\AA}}$, defined as the flux density ratio at the rest-frame 6 cm and that at 2500 \angstrom\ for the subset of those with available radio observations (see discussion below in \S\ref{subsec:radio_fraction}).

\subsection{SED Data from Archival Observations}

We queried the archival SED data for every source in the G15 and C16 catalogs using the Vizier tool \footnote{http://vizier.u-strasbg.fr/vizier/sed/} within 3$''$. This results in a combined sample of 138 periodic quasars with available photometry in more than 5 bands. We adopt measurements from large systematic surveys to focus on a more homogeneous dataset. These include the Galaxy Evolution Explorer \citep[GALEX;][]{Martin2005}, the Sloan Digital Sky Survey \citep[SDSS;][]{York2000}, the Two Micron All Sky Survey \citep[2MASS;][]{skrutskie06}, the Wide-field Infrared Survey \citep[WISE;][]{Wright2010}, the NRAO VLA Sky Survey \citep[NVSS;][]{Condon1998}, and the Faint Images of the Radio Sky at Twenty centimeters (FIRST) survey \citep{becker95}.

 \begin{figure*}
 \centering
  \hspace*{-10mm}
  \includegraphics[width=200mm]{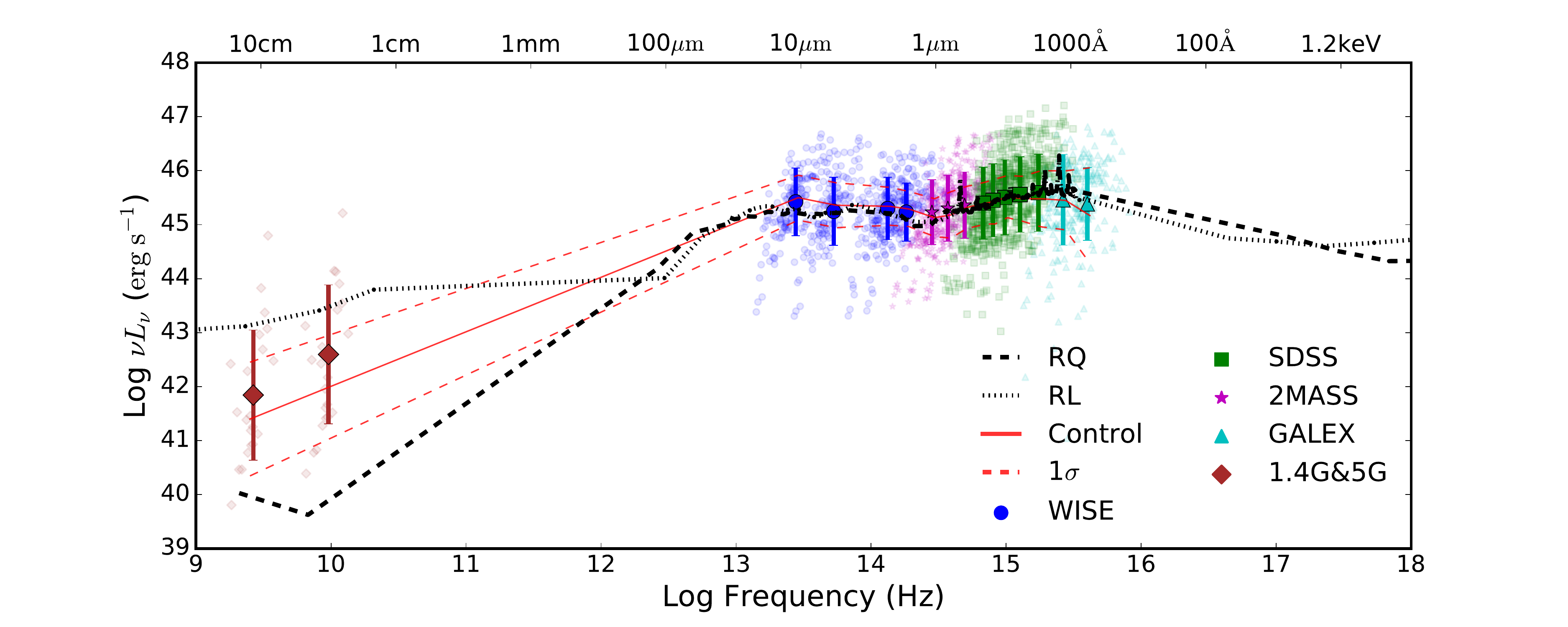}
    \caption{Rest-frame composite SED for candidate periodic quasars. Big symbols with error bars represent the geometric mean of $\nu L_{\nu}$ and the 1$\sigma$ dispersion, whereas small symbols denote individual candidate periodic quasars. For comparison, the solid and dashed red curves show the mean composite SED for control quasars and their 1$\sigma$ dispersion. Also shown for context are the average SEDs of radio-quiet (RQ; black dashed) and radio-loud (RL; black dotted) optically bright, non-blazar quasars from \citet{Shang2011}. All SEDs have been normalized to SDSS $i$ band.
 }
  \label{fig:sed}
\end{figure*}

When multi-epoch photometries are available, we take the mean value to quantify the average SED. When photometries are unavailable in some bands caused either by non-detection or by not being covered in the surveys, we repair the gaps in the UV-optical-IR SED following \citet{Richards2006}. This affects $<$5\% of our sample without SDSS or WISE measurements, and $\sim$30\% without 2MASS or GALEX measurements. The gaps are repaired by extrapolating the flux density in the nearest neighbouring band assuming the average SED of optically bright, non-blazar quasars \citep[including both radio-quiet and radio-loud objects;][]{Shang2011}. 

To calculate the radio loudness parameter $R_{6\,{\rm cm}/2500\,\mbox{\AA}}$, we adopt the 1.4 and 5 GHz data to calculate the flux density at the rest-frame 6 cm. For the 1.4 GHz detected sources without 5 GHz data, we extrapolate the SED assuming a radio spectral index $\alpha = -0.5$, where $F_{\nu} \propto \nu^{\alpha}$. 

All SEDs have been shifted to the quasar's rest frame and normalized to a small window (50 \angstrom\ around 7625 \angstrom ) close to the SDSS $i$ band which is chosen to be relatively free of strong emission lines. Galactic extinctions have been applied using the extinction map of \citet{schlegel98} assuming the reddening law of \citet{cardelli89}. 

%\section{The Properties of Spectral Energy Distributions}

%%%%%%%%%%%%%%%%%%%%%%%%%%%%%%%%%%%%%%%%%%%%%%%%%%%%%%%%%%%%%%%%%%%%%%%%%%%%%%%%%%%%%%%%%%%%%%%%
\section{Analysis and Results}\label{sec:sed}

To explore possible circumbinary accretion signatures in the SEDs of candidate periodic quasars, we first construct their mean SED and compare with that of the control quasars to look for any systematic difference between the two populations (\S \ref{subsec:mean_sed} \& \S \ref{subsec:radio_fraction}). We then inspect the SEDs of individual candidate periodic quasars to look for evidence of any significant deviations from typical quasar SEDs based on a colour selection and identify a sample of potential ``outliers'' with abnormally red SEDs (\S \ref{subsec:outlier}).  

\subsection{The Mean IR-Optical-UV SED of Candidate Periodic Quasars Is Similar to That of Control Quasars}\label{subsec:mean_sed}

Figure \ref{fig:sed} shows the composite SED of the sample of 138 candidate periodic quasars. We show both the mean value (large filled symbols) the 1-$\sigma$ dispersion (error bars) of the sample, as well as the individual objects (small ones). Also shown for comparison is the composite SED of the control sample of ordinary quasars that are matched to have the same redshift and $i$-band absolute magnitude distributions to those of the candidate periodic quasars (with the mean value shown in solid red and the 1-$\sigma$ ranges shown in dashed red), as well as the average SED of optically bright, non-blazar quasars of \citet{Shang2011}. 

The mean IR-Optical-UV SED of candidate periodic quasars is similar to that of both the control sample of ordinary quasars (in terms of both the mean value and the 1-$\sigma$ dispersion) and the \citet{Shang2011} sample of optically bright, non-blazar quasars. There is no evidence for any systematic difference or abnormal features, such as a notch or a cutoff. Our results do not change when we remove those objects with repaired SED gaps.

\subsection{Candidate Periodically-Variable Quasars Have A Higher Blazar Fraction than That of Control Quasars}\label{subsec:radio_fraction}

Figure \ref{fig:sed} also shows the composite radio SED of the radio-detected subset of 22 objects out of the 138 candidate periodic quasars. The average radio SED is similar to that of the radio-detected subset of 200 objects in the control sample of ordinary quasars, and is in between the radio-quiet (RQ; black dashed) and radio-loud (RL; black dotted) sub-populations of the \citet{Shang2011} sample of optically bright, non-blazar quasars. 

As shown in Figure \ref{fig:para}, while the radio-detected fraction of the candidate periodic quasars (22 out of 138, or $\sim$16$\pm$3\% where the uncertainty represents 1$\sigma$ Poisson error) is consistent with that of control quasars (200 out of 1380, or $\sim$14$\pm$1\%), the radio-loud (i.e., $R>10$) fraction (19 out of 138, or $\sim$14$\pm$3\%) is higher than control quasars (120 out of 1380, or $\sim$9$\pm$1\%) at the $\sim$2.5$\sigma$ significance level on average. In particular, the fraction of extremely radio-loud population with $R>100$ (13 out of 138, or $\sim$9$\pm$3\%), e.g., blazars, is tentatively higher than that of control quasars (50 out of 1380, or $\sim$4$\pm$1\%) at the $\sim$2.5$\sigma$ level on average.

\subsection{SED Properties of Individual Candidates: Identifying ``Outliers'' by Selecting Red Quasars}\label{subsec:outlier}

\begin{figure*}
 \centering
   \hspace*{-20mm}
  \includegraphics[width=220mm]{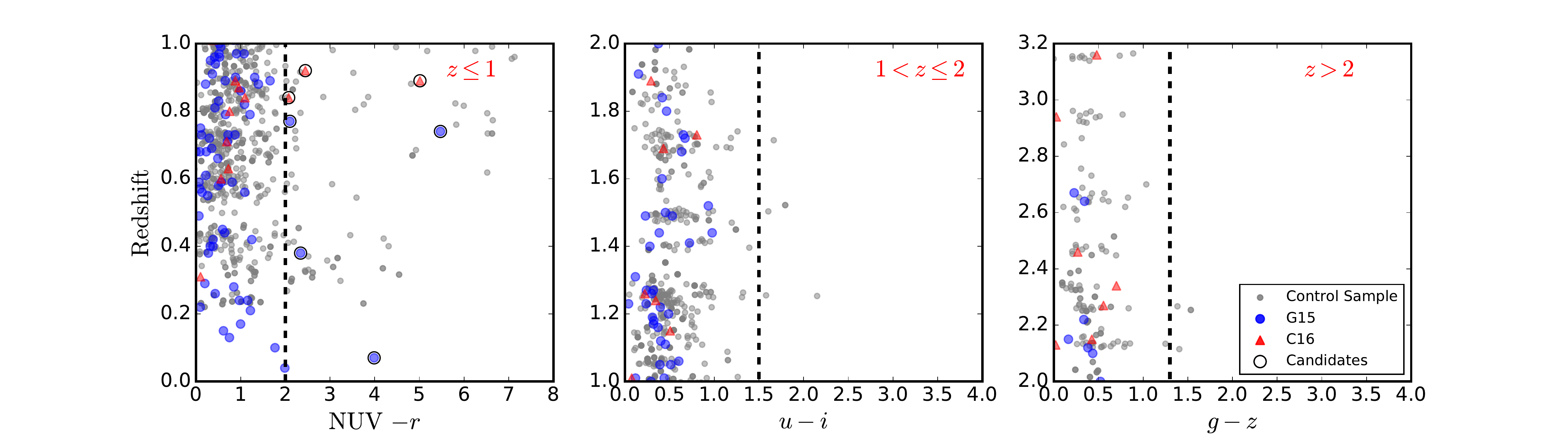}
    \caption{Colour selection of candidate periodic quasars with abnormally red SEDs using the criterion given by Equation \ref{eq:color}. Panels show the three redshift ranges in which different colour cuts are applied (dashed lines).   
    Big black open circles represent candidate periodic quasars that show abnormally red colours. Small grey dots denote control quasars. The red SED fraction of candidate periodic quasars ($\sim$5$\pm$2\%) is consistent with that in control quasars ($\sim$6$\pm$1\%). See \S \ref{subsec:outlier} for details.    
}
  \label{fig:color-z}
\end{figure*}

Some subtle, abnormal features may have been smoothed out due to the averaging effect in producing the composite SEDs. To investigate this possibility, we now inspect more closely the SED properties of individual candidate periodic quasars.

As discussed in \S \ref{sec:sed} and illustrated in Figure \ref{fig:theory}, both the cutoff due to a central cavity and the notch produced by minidiscs in circumbinary accretion discs will cause the a flux deficit in the bluer part of the IR-optical-UV quasar spectrum. Therefore, we can select possible ``outliers'' by identifying abnormally red quasars. We define an empirical colour criterion to select abnormally red quasars, which is given by:
\begin{equation}\label{eq:color}
\begin{split}
{\rm NUV} - r &> 2, {\rm when}\, z \leq 1, \\
u - i &> 1.5, {\rm when}\, 1 < z \leq 2, \\
g - z &> 1.3, {\rm when}\, 2 < z \lesssim 3,
\end{split}
\end{equation}
where the central wavelengths for GALEX NUV and SDSS $ugriz$ are 2329 \AA , 3543 \AA , 4770 \AA , 6231 \AA , 7625 \AA , and 9134 \AA, respectively. For $z>1$ quasars, the two colour cuts are estimated assuming a rest-frame reddened power-law spectrum with $f_{\lambda}\propto \lambda^{0.9}$ but k-corrected to higher redshifts. The NUV $- r$ colour index of the deepest notch profile (a factor of 3 fainter than the single BH case in \S \ref{sec:pred}) corresponds to $\lesssim$ 2 within the redshift range of 0 $\sim$ 1, while the cutoff profile is with respect to the NUV $-r$ colour in the redshift range of $z\sim$ 2--3.5. Therefore, by design our colour selection is sensitive to all cutoff cases and some deepest notch scenarios.

%If the redshift is larger than 1, the NUV photometric point will exceed the $\rm Ly \alpha$ and fall into the Lyman forest, as well as $u$ and $g$ bands for higher redshift.

Figure \ref{fig:color-z} displays the NUV$-r$, $u-i$ and $g-z$ colour vs.redshift for all the candidate periodic quasars in our SED sample. It illustrates the colour selection for objects at different redshift regimes. Seven red quasars satisfy Equation \ref{eq:color} (shown as black open circles in Figure \ref{fig:color-z}, including four objects from G15 and three objects from C16), all of which are at $z<1$. We discuss two individual quasars in detail in Appendix \ref{sec:appendix}.

%None is picked out by the latter two criteria since most our quasars are at low redshift. 
%XL: this statement makes little sense, however, because about half of the sample is at z>1.

Figure \ref{fig:cand} shows the individual SEDs and SDSS spectra for the seven red quasars. Also shown for comparison are SEDs for control samples of ordinary quasars that are individually drawn to match the redshift and $i$-band absolute magnitude for each particular quasar. Compared to the control sample, all the red quasars show significant flux deficits in the bluer part of the SED by selection. While being a relatively rare population, the fraction of red quasars among the parent sample of candidate periodic quasars (7 out of 138, or $\sim$5$\pm$2\%) is consistent with that in the control sample of ordinary quasars (89 out of 1380, or $\sim$6$\pm$1\%).

% table 1
\begin{table*}
	\centering
	\caption{Properties of candidate periodic quasars with red colours.}
	\label{table:table1}
	\begin{adjustbox}{width=\textwidth}
	\begin{tabular}{lcccccccccccc} % 
	        \hline
		\hline
& & & $ \rm Log \it M_{\rm BH}$ & $ L_{\rm bol}$ & a& Period & $T_{\rm cutoff}$ & $\lambda_{\rm cutoff}$ & $T_{0}$ & $\lambda_{\rm notch\;range}$& $\lambda_{\rm notch}$ & $\rm Depth$ \\
ID & SDSS Designation & Redshift & ($M_{\odot}$) & ($\rm erg\, s^{-1}$) & ($10^{-3}$pc ($\rm R_{g}$)) & (days) & (K) & (\AA)  & (K) & (\AA) & (\AA)  & (dex) \\
(1)     & (2)  & (3)    & (4)     &   (5)   &       (6)     &    (7)     &    (8)        &     (9)       &      (10)   &    (11)      &   (12)       & (13) \\
\hline
1 & J072908.71$+$400836.6 &0.074& 7.74$\pm$0.32 & 44.92 & 1.0 (374) & 1612 (2$\sigma$) & $2500$ & $11200$& $4300$  &400 - 6700& $1700$ & 1.6 \\%0.12
2 & J080648.65$+$184037.0 &0.745& 7.99$\pm$0.27 & 45.10 & 3.0 (630) & 892 (2$\sigma$) & $1700$   & $17400$& $2800$  &700 - 10400& $2600$ & 1.8\\ %0.10
3 & J081617.73$+$293639.6 &0.769& 9.77$\pm$0.33 & 46.15 & 13 (46) & 1162 (2$\sigma$) & $7300$ & $3900$  & $12400$&200 - 2300& $600$ & 1.5\\ %0.02
4 & J134553.57$+$334336.0 &0.885& 8.73$\pm$0.31 & 45.51 & 5.6 (214) & 797 (2$\sigma$)& $3000$  & $9600$   & $5100$  &400 - 5700& $1400$ &2.0 \\   %0.04
5 & J153636.22$+$044127.0 &0.389& 8.82$\pm$0.28 & 46.14 & 7.0 (218) & 1111(2$\sigma$) & $4000$ &$7100$   & $6800$   &300 - 4300& $1100$ &1.1\\ %0.16
6 & J214036.77$+$005210.1 &0.922& 8.50$\pm$0.31 & 45.73 & 2.5 (162) & 316 (1$\sigma$) & $4800$  &$6000$    & $8100$   &200 - 3600& $900$ & 2.0\\ 
7 & J232135.73$+$173916.5 &0.842& 8.68$\pm$0.31 & 45.46 & 3.4 (146) & 337 (1$\sigma$)& $4000$   &$7200$   & $6800$   &300 - 4300& $1100$ & 1.2\\ 
\hline
\multicolumn{13}{l}{Column 1: Object ID as labeled on Figure \ref{fig:wise_color}.}\\
\multicolumn{13}{l}{Column 2: SDSS names with J2000 coordinates given in the form of ``hhmmss.ss+ddmmss.s''.}\\
\multicolumn{13}{l}{Column 3: Systemic redshift from G15 and C16.}\\
\multicolumn{13}{l}{Column 4 \& 5: Total virial black hole mass (based on the width of broad emission lines in quasar spectra)} and bolometric luminosity from \citet{shen11} and \citet{Kozlowski2017}.\\
\multicolumn{13}{l}{~~~~~~~~~~~~~~~~~~$L_{\rm bol}$ is derived from the monochromatic luminosity at 5100 \AA\ assuming the bolometric correction 9.26 from \citet{richards06}.}\\
\multicolumn{13}{l}{Column 6: Semi-major axis estimate from G15 and C16 based on the Kepler's law.} \\
\multicolumn{13}{l}{Column 7: Periodicity from G15 and C16 and significance level of the periodicity from our new estimates (\S \ref{subsec:false_positive}).}\\
\multicolumn{13}{l}{Columns 8 \& 9: Expected cutoff temperature and the corresponding wavelength at the inner edge of the circumbinary disc (\S \ref{sec:pred}).}\\
\multicolumn{13}{l}{Columns 10--12: Expected notch temperature and the corresponding wavelength range and wavelength of the deepest notch (\S \ref{sec:pred}).}\\
\multicolumn{13}{l}{Column 13: Highest flux deficit observed, defined as the largest difference between the observed SED and the mean SED of control quasars.}\\
\end{tabular}
\end{adjustbox}
\end{table*}

% table 2
\begin{table}
	\centering
	%\tabletypesize{\scriptsize}
	\caption{Best-fit dust extinction model parameters for candidate periodic quasars with red colours.}
	\label{table:table2}
	%\begin{adjustbox}{width=\textwidth}
	\begin{tabular}{lcccc} % 
	        \hline
		\hline
 & $A_V$&  &  & \\
ID & (mag) & $c_1$& $c_2$ &$\chi^2_{\nu}$\\
(1) & (2) & (3) & (4) & (5) \\
\hline
1 & 1.08$\pm$0.13 & $-$4.61$\pm$0.47 & 3.61$\pm$0.10 & 1.22 \\
2 & 0.16$\pm$0.07 & $-$4.58$\pm$0.50 & 1.02$\pm$0.47 & 1.12 \\
3 & 0.11$\pm$0.03 & $-$5.91$\pm$0.37 & 2.37$\pm$0.27 & 1.08 \\
4 & 0.40$\pm$0.05 & $-$4.72$\pm$0.39 & 2.42$\pm$0.46 & 1.21 \\
5 & 0.16$\pm$0.11 & $-$5.62$\pm$0.21 & 3.46$\pm$0.40 & 0.99 \\
6 & 0.27$\pm$0.04 & $-$6.00$\pm$0.27 & 2.35$\pm$0.22 & 0.93 \\
7 & 0.35$\pm$0.12 & $-$10.0$\pm$0.82 & 4.02$\pm$0.33 & 1.37 \\
\hline
\multicolumn{5}{l}{Column 1: Object ID as listed in Table \ref{table:table1}.}\\
\multicolumn{5}{l}{Columns 2--4: Best-fit value and 1$\sigma$ error of the free parameters in the}\\ 
\multicolumn{5}{l}{~~~~~~~~~~~~~~~extinction curve model given by Equation \ref{eq:fm}.}\\
\multicolumn{5}{l}{Column 5: $\chi^2$ per degree of freedom in the FM model fit.}\\
\end{tabular}
%\end{adjustbox}
\end{table}

%%%%%%%%%%%%%%%%%%%%%%%%%%%%%%%%%%%%%%%%%%%%%%%%%%%%%%%%%%%%%%%%%%%%%%%%%%%%%%%%%%%%%%%%%%%%%%%%
\section{Discussion}\label{sec:dis}

\subsection{Comparison with Circumbinary Accretion disc Models for Candidate Periodic Quasars with Red SEDs}\label{subsec:sed_location}

As described in \S\ref{sec:pred}, we can estimate the wavelengths of the SED cutoff or notch using the characteristic temperatures $T_{{\rm cutoff}}$ and $T_{{\rm notch}}$ using Equation \ref{eq:T}. We adopt the virial BH mass estimate $M_{{\rm BH}}$ of \citet{shen11} and \citet{Kozlowski2017} based on the width of the broad \hb\ emission line. The accretion rate is $\dot{m}= \dot{M}/\dot{M_{{\rm Edd}}}= \frac{L_{{\rm bol}}/(\eta c^2)}{L_{{\rm Edd}}/(\eta c^2)} = L_{{\rm bol}}/L_{{\rm Edd}}$, where $L_{\rm Edd}= 1.26\times 10^{38} M_{{\rm BH}}/M_{\odot}$ $\rm erg\, s^{-1}$ \citep[e.g.,][]{Foord2017}. We estimate $L_{\rm bol}$ from the monochromatic luminosity at 5100 \AA\ assuming the bolometric correction of \citet{Richards2006}. We assume that the radiative efficiency of the accretion disc is $\eta= 0.1$. We estimate the binary's semimajor axis using the candidate periods reported by G15 and C16 assuming a circular orbit and that the binary orbital period is the same as the period in the optical light curve. Table \ref{table:table1} lists the resulting characteristic temperatures and the corresponding wavelengths of the expected SED cutoff and notch and the depth of the deepest notch.

\begin{figure*}
 \centering
  \vspace*{-1cm}
  \hspace*{0cm}
  \includegraphics[width=85mm,height=220mm]{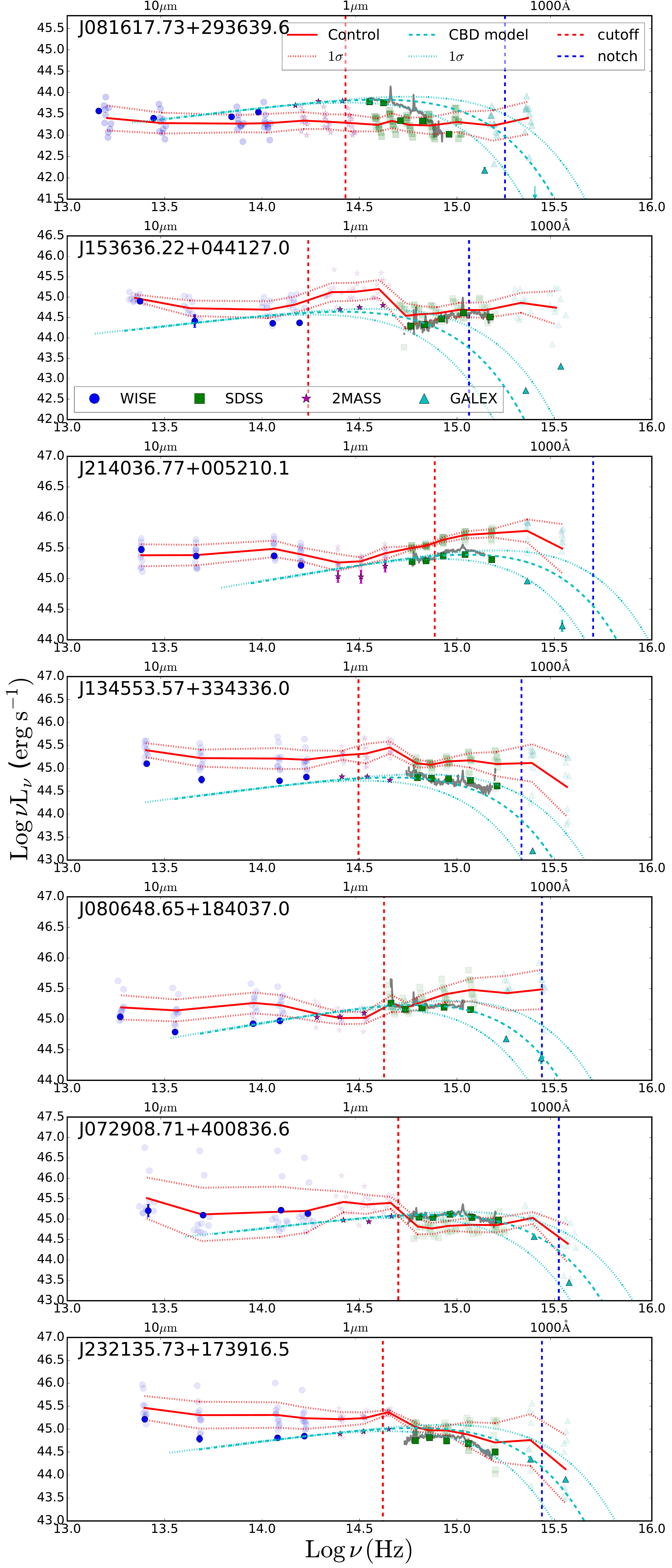}
  \includegraphics[width=85mm,height=220mm]{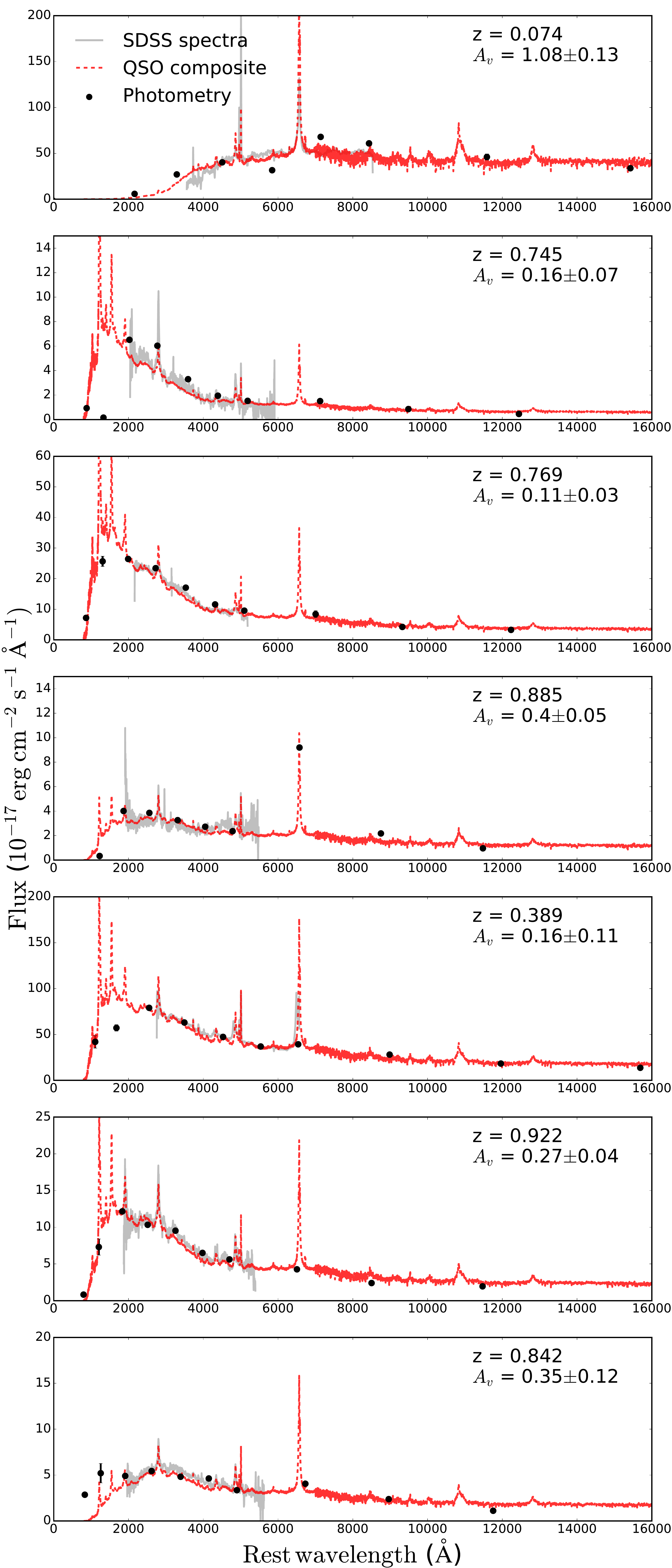}
    \caption{SEDs (left column) and SDSS spectra (right column) for seven candidate periodic quasars with red colours selected by the colour criterion defined in Equation \ref{eq:color}. In the left column, solid symbols show SEDs for candidate quasars whereas lighter symbols represent control quasars for each object matched in redshift and luminosity. Solid and dotted red curves show the mean SED of control quasars and their 1$\sigma$ dispersion. Blue and red dashed lines mark the expected locations of SED cutoff and notch (\S \ref{sec:pred}). The dashed and dotted cyan curves show the CBD (cutoff) model and its 1 $\sigma$ ranges considering the $\sim$ 0.5 dex systematic scatter in the virial BH mass estimates \citep[e.g.,][]{Shen2011}. The models have been normalized to the observed SEDs at the cutoff wavelength. In the right column, grey solid curves show the SDSS spectra for candidate periodic quasars. Red dashed curves represent the best-fit model constructed using a composite quasar SED \citep{VandenBerk2001,Glikman2006} reddened by a extinction curve model given in Equation \ref{eq:fm} following the FM formalism \citep{Fitzpatrick1986,Zafar2011,Zafar2015}. The best-fit V-band dust extinction $A_{v}$ ranges from $\sim$ 0.1 to 1.1 mag. See \S \ref{subsec:dust} for details.
}
  \label{fig:cand}
\end{figure*}

As shown in Figure \ref{fig:cand}, the expected SED cutoff wavelength is close to the location where the SEDs of BSBH candidates start to be systematically redder than those of the control quasars, which verifies our colour selection. Considering the colour selection criterion (\S \ref{subsec:outlier}), six out of the seven candidate periodic quasars show SEDs that are broadly consistent with predictions under the cutoff scenario accounting for a $\sim$ 0.5 dex systematic scatter in the virial BH mass estimates \citep[e.g.,][]{Shen2011}. The other object (i.e., J153636.22+044127.0) is tentatively ruled out, given the inconsistency between the observed SED and the CBD model. Although the SEDs of some candidates deviate from the CBD model by 1$\sigma$, they are still broadly consistent with the cutoff scenario considering model parameter uncertainties.
On the other hand, the notch scenario is less likely for these candidates with red SEDs, because: (1). the SEDs do not seem to be turning up blueward of the notch locations, although the available SED data cannot rule out this possibility, and (2) assuming that the notch locations are bluer than the available SED data, the implied highest flux deficit in the deepest portion of the SED notch is typically beyond 1.0 dex (Table \ref{table:table1}), which is a factor of 2 of the theoretical prediction (at most a factor of $\sim$3, i.e., 0.5 dex, \citealt{Roedig2014}). 

In summary, the SEDs of the colour-selected periodic candidates are broadly consistent with theoretical predictions from the circumbinary accretion disc models with cutoffs due to a central cavity, where the inner region of the circumbinary disc is almost emptied by the secondary BH. On the other hand, the minidisc scenario (i.e., substantial accretion onto one or both BHs with their own minidiscs) is likely disfavored, although the available SED data cannot rule out this possibility entirely given the limited coverage and possible uncertainties in the model parameters. One possible caveat is that there may be candidate periodic quasars that would satisfy the minidisc scenario (i.e., with a notch in the SED) if we relax the colour criterion defined in Equation \ref{eq:color}. We have found no convincing candidate for such a shallower but wide enough notch feature, however, by examining the individual SEDs for all candidate periodic quasars.
A potential caveat is that wide-separation binaries, whose secondary may open a wide gap in the NIR or mid-IR regime, may be missed by our colour selection criteria. Nevertheless, we have examined all the individual candidate SEDs and found no convincing candidate with such a NIR/mid-IR cutoff or notch.

\subsection{Alternative Explanation for the SED Outliers in Candidate Periodic Quasars: Dust Reddening}\label{subsec:dust}

Alternatively, the unusually red quasar colours in the SED outliers may be due to reddening by dust either in the immediate surroundings of the accretion discs and/or in the quasar host galaxies \citep[e.g.,][]{Leighly2016}. To explain the observed outlier SEDs with dust reddening, we assume a model using a composite quasar SED reddened by an extinction curve model. For the composite quasar SED, we adopt the optical/UV composite of $\sim$2000 SDSS quasars from \citet{VandenBerk2001} for $\lambda<$7000 \AA\ and the NIR composite of 27 quasars from \citet{Glikman2006} for $\lambda\ge$7000 \AA . To model the extinction curve, we follow the \citet[][which we refer to as ``FM'' below]{Fitzpatrick1986} formalism \citep[see also][]{Zafar2011,Zafar2015}, which is given by
\begin{equation}\label{eq:fm}
A_\lambda = A_V \left[\frac{1}{R_V}k(\lambda-V) + 1\right],
\end{equation}
where 
\[k(\lambda-V) = \left\{ 
\begin{array}{l l}
  c_1+c_2x+c_3D, & \quad \mbox{when $x\leq c_5$ }\\
  c_1+c_2x+c_3D+c_4(x-c_5)^2, & \quad \mbox{when $x >c_5$ }\\ \end{array} \right. \]
in which $x\equiv \lambda^{-1}$, $A_{V}$ is the V-band dust extinction, and $R_V$ is the total-to-selective extinction given by $R_V\equiv A_V/E(B-V)$. We adopt a classical Small Magellanic Cloud type extinction curve which is commonly used to model reddened quasar spectra \citep[e.g.,][]{Richards2003,Hopkins2004,Glikman2012,Zafar2015}, i.e., setting $R_V=2.93$ \citep{Pei1992}. The FM formalism consists of two components: (1). a UV linear component modeled by the parameters $c_1$ (intercept) and $c_2$ (slope) and the far-UV curvature modeled by the parameters $c_4$ and $c_5$, and (2). a Drude function $D$ that describes the 2175 \AA\ bump; for simplicity, we assume no 2175 \AA\ bump in our model, i.e., setting $c_3=0$ in Equation \ref{eq:fm}. As the archival SED data does not provide enough far-UV coverage to fit for $c_4$ and $c_5$, we fix them to be the average values of known reddened quasars, i.e., assuming $c_4=0.62$ and $c_5=5.9$ \citep{Zafar2015}. Our final extinction curve model contains three free parameters, i.e., $A_V$, $c_1$, and $c_2$.

We fit the dust extinction model in Equation \ref{eq:fm} to the GALEX-SDSS-2MASS part of the SED data with the mpfit package \citep{Markwardt2009} using a least-$\chi^2$ minimization algorithm.  We have normalized the data to the $K_s$ band which is the least affected by dust. We have scaled the SDSS spectra based on the $i$ band since it is free from strong emission lines for all objects considered. For data points without error measurements (e.g., FUV and NUV derived from force photometry), we assume a fiducial error of 15\% of the local flux density. We assume the fitting ranges of the three parameters to be $A_{V}\in$[0, 2] mag, $c_1\in$ [$-$10.0, 2.0], and $c_2\in$ [0.15, 1.45], respectively.

The right panel of Figure \ref{fig:cand} shows the best-fit dust extinction model for each outlier SED candidate. In general, the model agrees with the data well considering measurement uncertainties and systematic errors due to quasar variability. Table \ref{table:table2} lists the best-fit value and 1$\sigma$ error for the three free parameters in the dust extinction model given by Equation \ref{eq:fm}. The errors are estimated using bootstrap re-sampling. The best-fit $A_V$ values range from 0.1 to 1.1 mag, which are reasonable for optical quasars \citep[e.g.,][]{Zafar2015}. Our results on the dust extinction modeling, together with the fact that the fraction of red, ``outlier'' quasars in candidate periodic quasars is similar to that in the control sample or ordinary quasars, suggest that dust reddening is more likely to explain the ``anomalous'' SEDs in candidate periodic quasars.

\subsection{Further Evidence for Dust Reddening in Candidate Periodic Quasars with Red SEDs}\label{subsec:wise}

 \begin{figure}
 \centering
  \vspace*{0cm}
  \hspace*{0cm}
  \includegraphics[width=85mm]{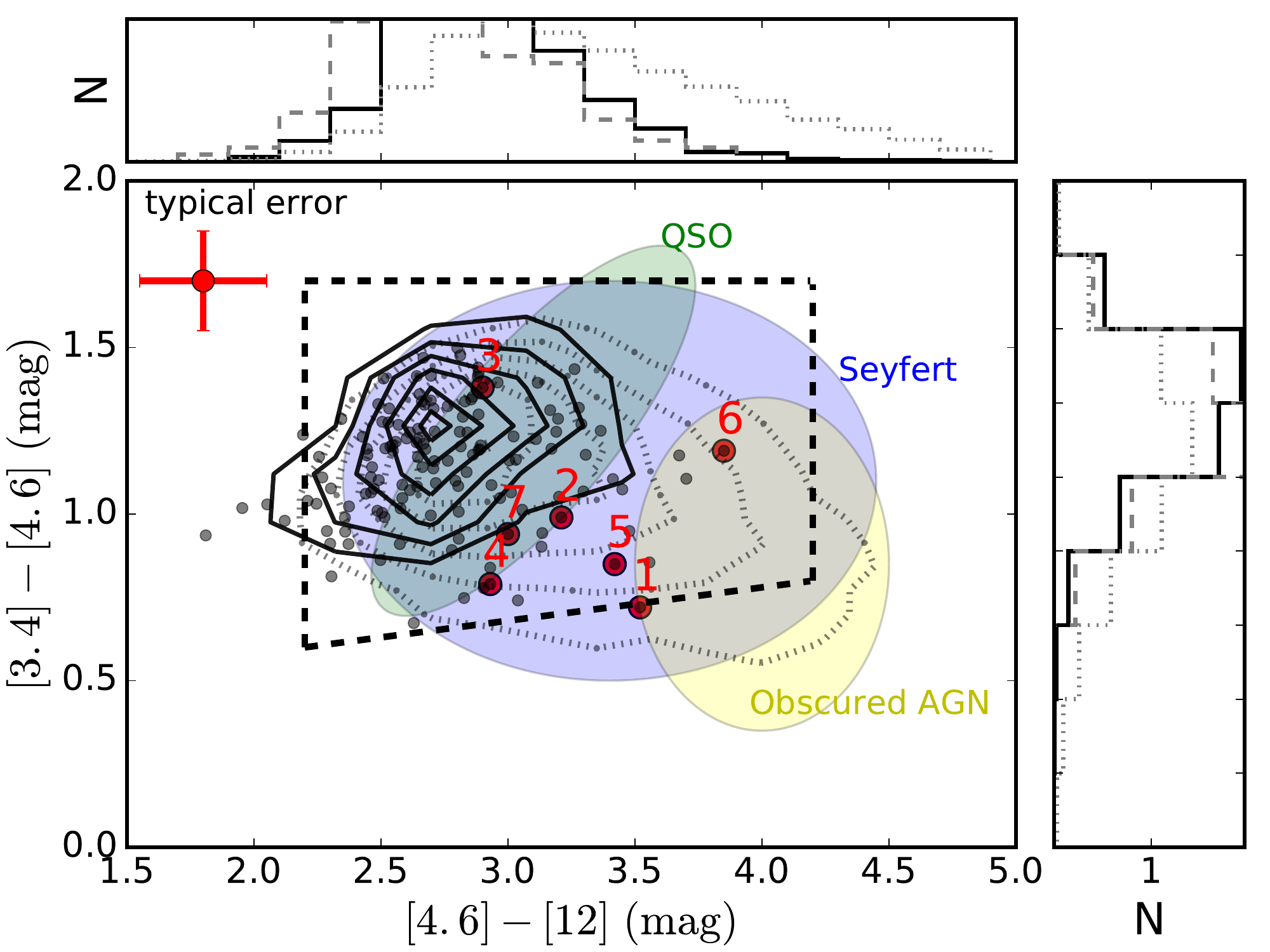}
    \caption{WISE colours for candidate periodic quasars. Grey dots show candidate periodic quasars of which those with red SEDs selected by Equation \ref{eq:color} are circled in red. Labeled are their IDs as listed in Table \ref{table:table1}. Shown for comparison are control quasars (black solid contours) and the parent sample of SDSS DR14 quasars (grey dotted contours). Also shown for context are the general AGN population (black dashed box) defined by \citet{Jarrett2011}, including QSOs (green ellipse), Seyferts (blue ellipse), and obscured AGN (yellow ellipse). Histograms show WISE colour distributions for candidate periodic quasars (grey dashed), control quasars (black solid), and SDSS DR14 quasars (grey dotted). See \S \ref{subsec:wise} for details.
}
  \label{fig:wise_color}
\end{figure}

We discuss further evidence for dust reddening in the candidate periodic quasars with abnormally red SEDs. Figure \ref{fig:wise_color} shows the WISE W2$-$W3 (i.e., [4.6]$-$[12] in mag) vs. W1$-$W2 (i.e., [3.4]$-$[4.6] in mag) colour-colour diagram for the sample of candidate periodic quasars with red SEDs. For context, the dashed box shows the region occupied by WISE AGNs empirically defined by \citet{Jarrett2011}, which include QSOs, Seyferts, and obscured AGNs. Also shown for comparison are the parent sample of 138 candidate periodic quasars, the control sample of ordinary quasars, as well as the SDSS DR14 quasars. The parent sample of 138 candidate periodic quasars has similar WISE colours to those of the control quasars. On the other hand, the subset candidate periodic quasars with abnormally red SEDs seems to be systematically skewed towards the obscured AGN population compared to control quasars and the SDSS DR14 quasars, consistent with the expectation from dust reddening.

\subsection{False Positives in Current Candidate Periodic Quasars from Few-Cycle Light Curves}\label{subsec:false_positive}

We discuss possible false positives in the current sample of candidate periodic quasars and their implications in the context of our SED results. \citet{Vaughan2016} has demonstrated that the candidate periodicity in the blazar PG1302$-$102 may be a false positive from random quasar variability \citep[see also][]{Liutt2018}. Considering that PG1302$-$102 was originally proposed as the best candidate in the G15 sample, it is possible that the candidate periodic quasars in the G15 and C16 samples are subject to similar uncertainties given the limitations of the observations (e.g., limited time baselines that cover only $\gtrsim$1.5 of the claimed cycles, uneven sampling and seasonal gaps in the cadence, and relatively low sensitivities).

In particular, to test the robustness of the seven candidate periodic quasars with red SEDs, we re-assess the significance of their periodicity based on the public CRTS or PTF light curves. We calculate the generalized Lomb-Scargle periodogram  \citep{Zechmeister2009} and construct a set of 50,000 simulated light curves for each quasars to more carefully assess the significance of any periodogram peak. We generate the simulated light curves assuming a damped random walk model \citep[DRW;][]{Kelly2009,Kozlowski2010,MacLeod2010,Butler2011,Ruan2012,Zu2013} for the stochastic red noise variability. We tailor the simulated light curves to each quasar by sampling the probability density function of the DRW model parameters as measured from the observed light curve. Following \citet{Vaughan2016}, we have also tested alternative models using the more general broken-power laws to model the power density of the observed quasar variability to verify that our results are robust against the DRW assumption, considering evidence for possible deviations from the DRW model on both short (inter-day) and long ($>$20 yr) timescales \citep[e.g.,][]{Mushotzky2011,Kasliwal2015,Guo2017,Smith2018}.

Table \ref{table:table1} lists the significance levels obtained from our re-analysis of the light curve periodicity for the seven candidates. While we were able to reproduce the reported periods, none of them exceeds the 3$\sigma$ significance level estimated assuming the DRW model; using alternative models to the DRW assumption would lower the statistical significance even further. In two of the seven objects, the candidate periodicity is only significant at the $\sim$1$\sigma$ level. 

In summary, our results are consistent with \citet{Vaughan2016} and suggest that the majority of the candidate periodic quasars reported based on few-cycle, noisy observations with uneven sampling and seasonal gaps may be false positives due to the stochastic, red noise quasar variability. In this scenario, it is unsurprising that the SEDs of candidate periodic quasars are similar to those of control quasars, considering that they would contain the same fraction of BSBHs, if any.

\subsection{Sampling Biases Driven by Variability Selection}\label{subsec:bias}

 \begin{figure}
 \centering
  \vspace*{-1cm}
  \hspace*{-3mm}
  \includegraphics[width=93mm]{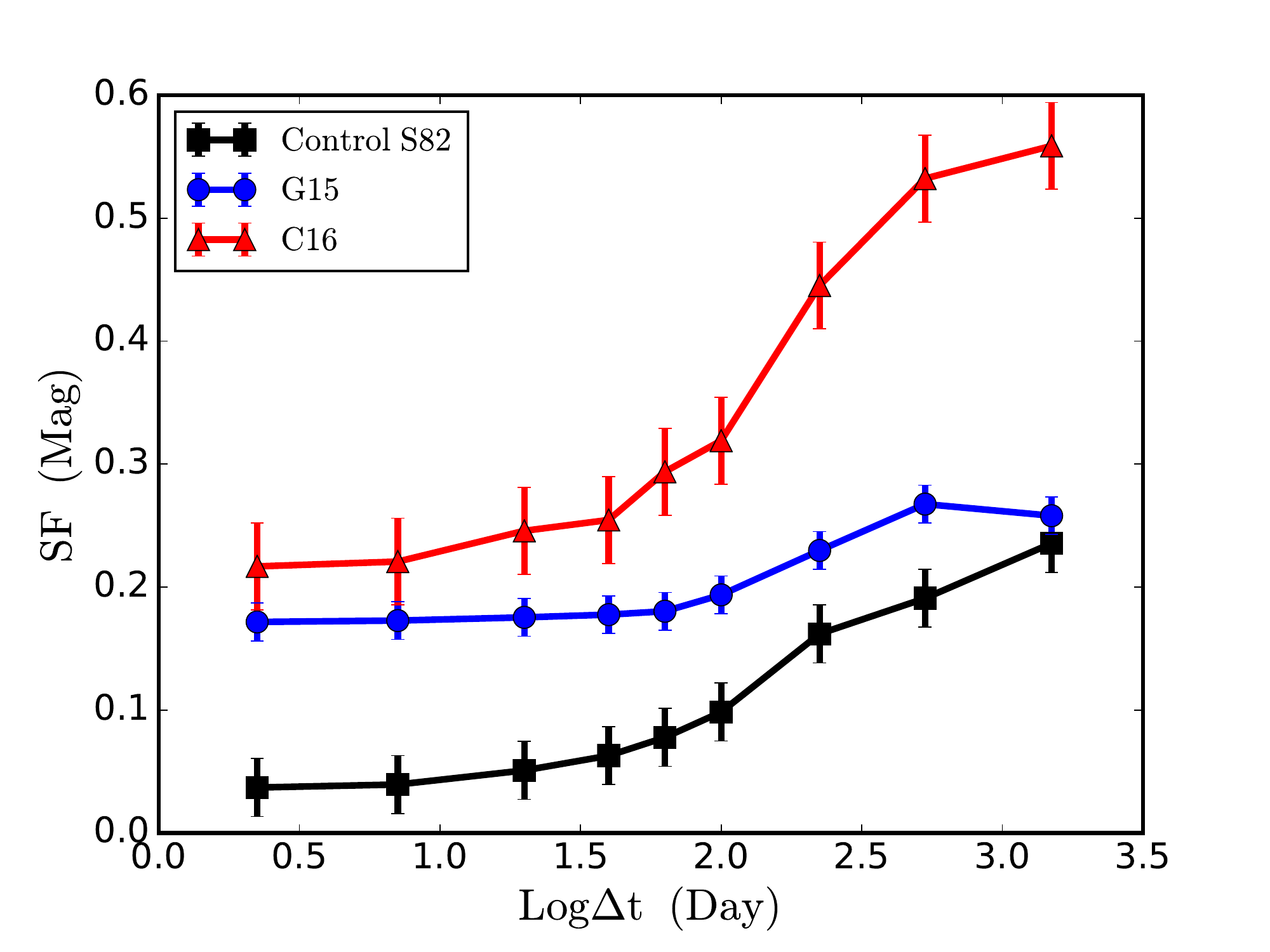}
  \includegraphics[width=85mm]{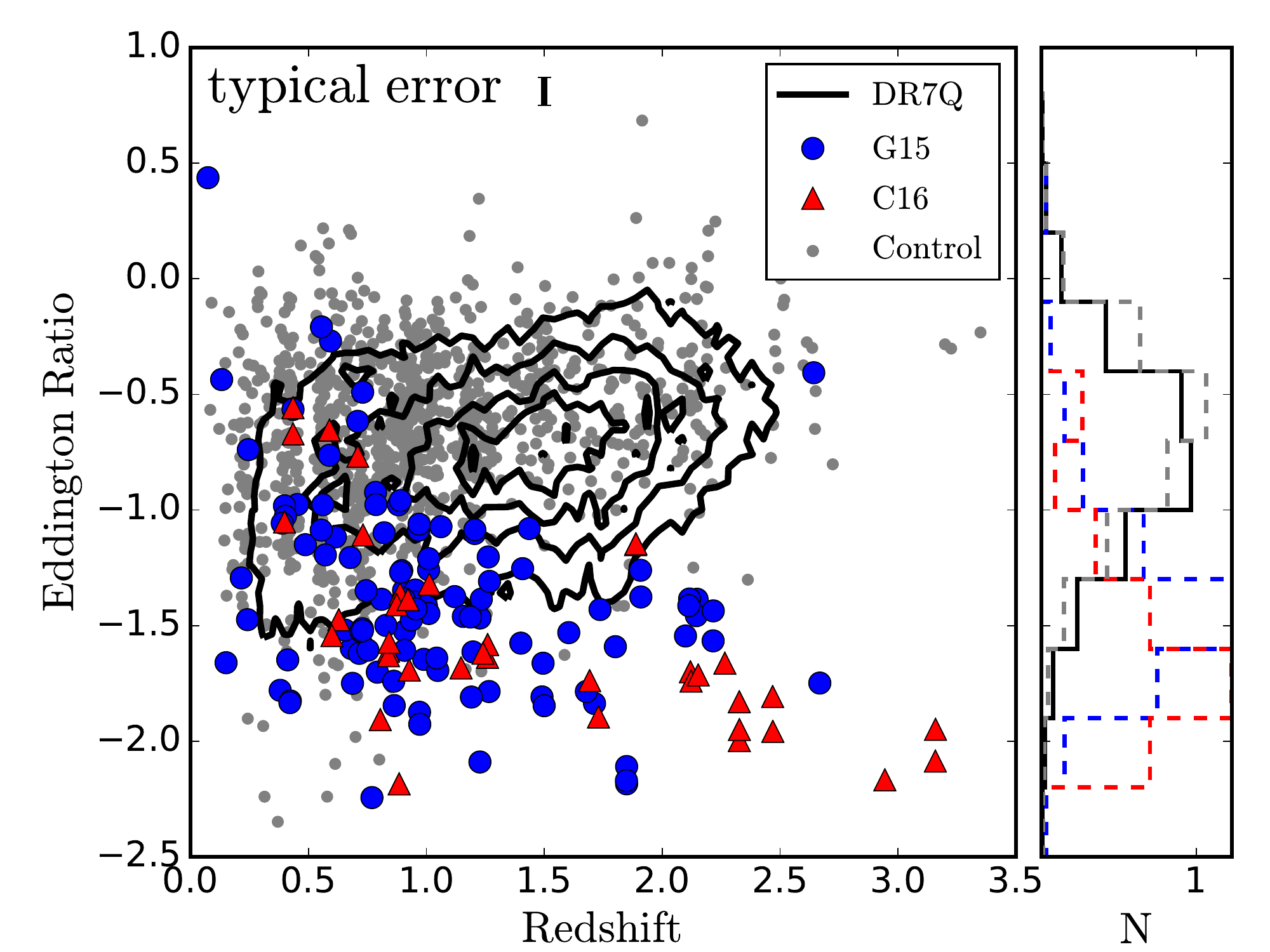}
    \caption{Upper panel: ensemble structure functions of candidate periodic quasars in the G15 (blue circles) and C16 (red triangles) samples compared against that of control quasars (black squares). Candidate periodic quasars are on average more variable than control quasars on all timescales. Low panel: Eddington ratio versus redshift for candidate periodic quasars (large blue circles and red triangles) compared against control quasars (small grey dots) and the parent sample of SDSS DR7 quasars in Stripe 82 (DR7Q; black contours). Histograms show Eddington ratio distributions for candidate periodic quasars from G15 (dashed blue) and C16 (red dashed), control quasars (grey dashed), and SDSS DR7 Stripe 82 quasars (solid black). See \S \ref{subsec:bias} for details.
}
  \label{fig:EDD}
\end{figure}

 \begin{figure}
 \centering
  \vspace*{-1cm}
  \hspace*{-3mm}
  \includegraphics[width=93mm]{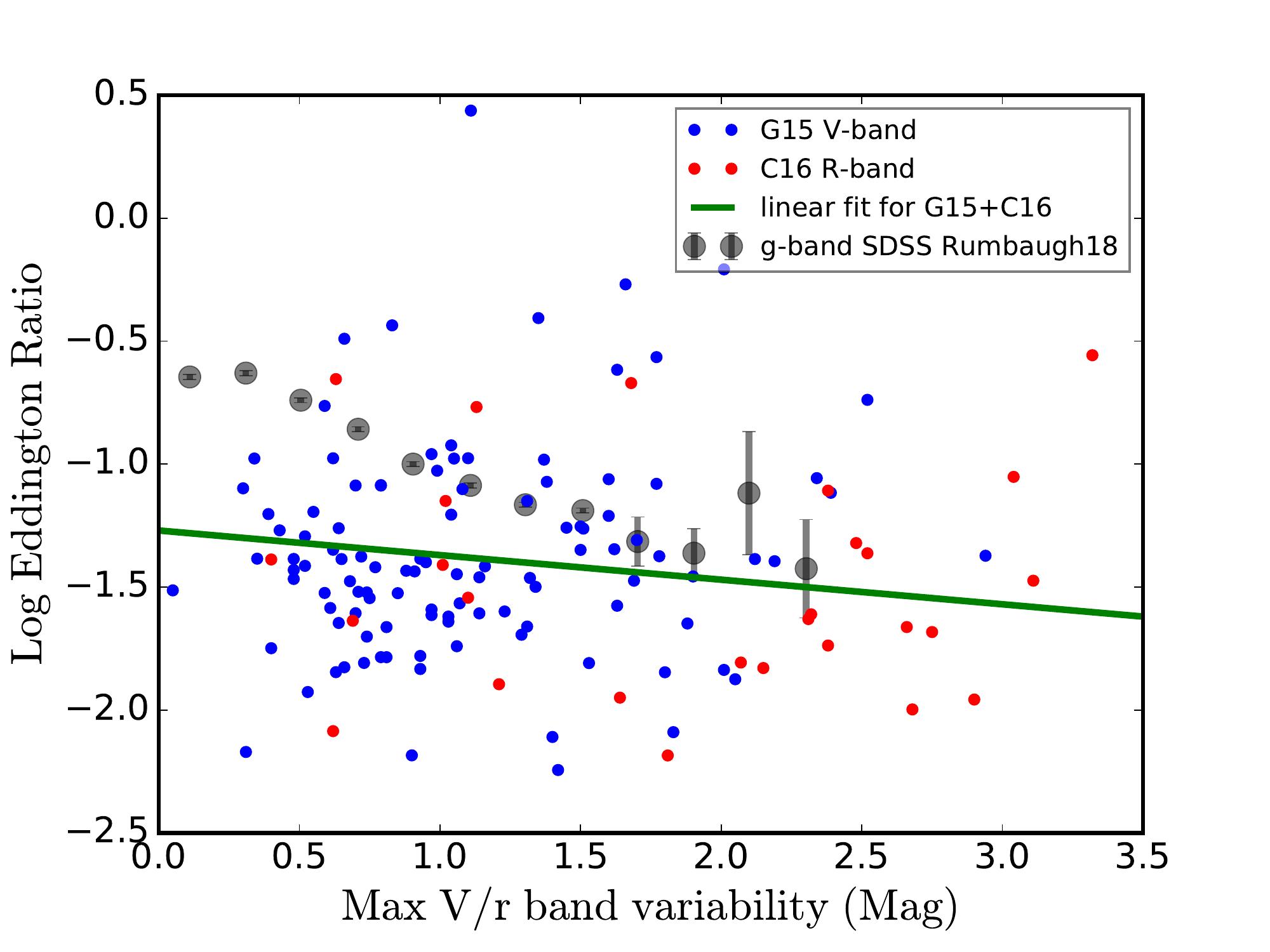}
    \caption{The anti-correlation between Eddington ratio and maximal optical variability amplitude observed in periodically-variable quasars (small coloured dots and the green line) compared against that in ordinary quasars (large grey filled circles). The grey filled circles denote the median Eddington ratio in each bin of maximal $g$-band variability amplitude for SDSS Stripe 82 quasars from \citet{Rumbaugh2017}. The green line represents the linear fit for the periodical quasar sample (G15+C16) in the $V$/PTF $R$ band.}
  \label{fig:comp}
\end{figure}

Figure \ref{fig:EDD} shows the ensemble structure function \citep[``SF'' for short; e.g.,][]{Sun2014,Kozlowski2016a,Sun2018a} for the candidate periodic quasars (in $V$ band for the G15 sample and in PTF $R$ band for the C16 sample). The SF describes aperiodic luminosity fluctuations by means of the rms variability as a function of the time difference between epochs. Also shown for comparison are the ensemble SFs (in SDSS $r$ band, which is similar to PTF $R$ band) for a control sample drawn from SDSS Stripe 82 quasars to match the redshift and absolute $i$-band luminosity distribution of candidate periodic quasars. Candidate periodic quasars are systematically more variable than the control sample over all the timescales considered (from days to a decade). The difference cannot be explained by the colour dependence of quasar variability given the small band differences (effective central wavelength of 547.7 nm for CRTS Johnson $V$ and 623.1 nm for SDSS $r$). 

The higher level of variability in candidate periodic quasars is largely driven by a selection bias considering that: (1). a significant periodicity is easier to detect in more variable quasars given the same measurement uncertainties, and (2). more variable quasars are likely to cause more false positives in periodicity searches based on few-cycle observations. 

More variable quasars are known to have systematically lower Eddington ratios \citep[e.g.,][see also \citealt{Guo2014}]{Rumbaugh2017}. Figure \ref{fig:EDD} shows that candidate periodic quasars have systematically lower Eddington ratios (by $\sim$1 dex on average) than control quasars, verifying the known anti-correlation between Eddington ratio and optical quasar variability. Among the candidate periodic quasar sample, the C16 subset is on average more variable than the G15 subset (upper panel in Figure \ref{fig:EDD}), because quasars from C16 have systematically lower luminosities (Figure \ref{fig:para}) and smaller Eddington ratios (lower panel in Figure \ref{fig:EDD}) than those from G15, where the colour-dependent variability amplitude difference is negligible (between CRTS $V$ and PTF $R$ bands).

Figure \ref{fig:comp} shows the maximal optical variability amplitude versus Eddington ratio for the periodically-variable quasars. An anti-correlation is observed where the green line represents the best-fit linear model for the periodic quasar sample. A similar anti-correlation has also been observed in normal SDSS quasars \citep{Rumbaugh2017} shown as grey filled circles. The slope in the anti-correlation is steeper in the normal quasar sample than that in the periodic quasar sample, which may be due to a combination of luminosity- and colour-dependent effects \citep[e.g.,][]{Macleod2012a}. The shallower slope observed in the periodic quasar sample suggests that the trend can be fully explained by that already observed in normal quasars. Nevertheless, we cannot rule out additional effects such as the emission being dominated by the sub-Eddington accretion of the primary BH whereas the super-Eddington accretion of the secondary BH is dim due to trapping of the radiation, which reduces the radiative efficiency under the binary scenario \citep[e.g.,][]{Farris2014}.

The tentatively higher blazar fraction found in candidate periodic quasars (\S \ref{subsec:radio_fraction}) may also be naturally explained as being driven by a variability-selection bias, given that blazars are also more variable in the optical \citep[e.g.,][]{Ruan2012}. On the other hand, if most candidate periodic quasars are robust, the higher blazar fraction could imply significant optical contamination from precessing radio jets \citep[e.g.,][]{Kudryavtseva2011,Ackermann2015a,Caproni2017}, because a sample selected to be of periodically-variable quasars would be more likely to contain jetted AGN, e.g., blazars, than a sample of normal quasars.

\begin{comment}
\begin{deluxetable*}{lccccccccc}
\tabletypesize{\scriptsize} 
\tablewidth{\textwidth} 
\tablecaption{Colour index of seven candidates
\label{tab:table3} 
} 
\tablehead{\colhead{} & \colhead{RA}& \colhead{DEC}  & \colhead{W1} & \colhead{W2}& \colhead{W3}  & \colhead{W1-W2}& \colhead{W2-W3} \\
\colhead{No.} & \colhead{deg} &\colhead{deg}& \colhead{mag} & \colhead{mag} & \colhead{mag}& \colhead{mag} & \colhead{mag} \\
\colhead{(1)} & \colhead{(2)} & \colhead{(3)} & \colhead{(4)} & \colhead{(5)} & \colhead{(6)} & \colhead{(7)}& \colhead{(8)}  } 
\startdata
1&112.286306& 40.143581&12.40&11.72& 8.88&0.72&3.52\\
2&121.702715& 18.676997&16.21&15.21&12.01&0.99&3.21\\
3&124.073949& 29.611059&14.16&12.78& 9.88&1.38&2.90\\
4&206.473385& 33.726632&15.52&14.73&11.80&0.79&2.93\\
5&234.150981& 4.690892 &12.90&12.05& 9.48&0.85&3.42\\
6&325.153222& 0.869478 &14.82&13.63&10.97&1.19&3.85\\
7&350.398886& 17.654576&15.37&14.53&11.53&0.94&3.00\\
\enddata
\tablecomments{Column 1: number.
}
\end{deluxetable*}
\end{comment}

%%%%%%%%%%%%%%%%%%%%%%%%%%%%%%%%%%%%%%%%%%%%%%%%%%%%%%%%%%%%%%%%%%%%%%%%%%%%%%%%%%%%%%%%%%%%%%%%
\section{Summary and Future Work}\label{sec:con}

Periodically-variable quasars have long been suggested as possible BSBH candidates, but alternative scenarios remain possible. As an independent and complementary test of the binary hypothesis, we have searched for evidence of a truncated or gapped circumbinary accretion disc by studying the SEDs of a sample of candidate periodic quasars. The sample combined the two largest candidate periodic quasar samples known from the CTRS and PTF surveys. Our work is motivated by recent circumbinary accretion disc simulations that predict abnormalities such as a cutoff or notch in the IR-optical-UV SED, depending on the model of circumbinary accretion and the evolutionary state of the system. To put our results into context, we have compared the SEDs of candidate periodic quasars against a control sample of ordinary quasars matched in redshift and luminosity. The work serves as a complementary test of the binary hypothesis for candidate periodic quasars. We summarize our main findings as follows.

\begin{enumerate}

\item The mean SED of candidate periodic quasars is similar to that of control quasars matched in redshift and luminosity (\S \ref{subsec:mean_sed}). Our results suggest that, if the candidate periodicity is robust, the SEDs of most circumbinary accretion discs may not be significantly different from accretion discs around single BHs, at least in the IR-optical-UR part, assuming the periodicity is indeed due to a binary. Alternatively, if most of the candidate periodic quasars are false positives (\S \ref{subsec:false_positive}), the similarity in the mean SED between candidate periodic quasars and control quasars will be unsurprising, considering that they would contain the same fraction of BSBHs. 

\item The fraction of radio loud quasars (i.e., with radio loudness $R>10$), and blazars (i.e., with $R>100$) in particular, is tentatively higher than that in the control sample (\S \ref{subsec:radio_fraction}). The higher radio-loud fraction, and a higher blazar fraction in particular, may be naturally explained as being driven by a variability-selection-induced sampling bias (\S \ref{subsec:bias}). On the other hand, if most periodic quasar candidates are robust, the higher blazar fraction could imply contamination from a processing radio jet.

\item Seven of 138 candidate periodic quasars show a significant cutoff in the IR-optical-UV SED (i.e., with abnormally red colours, \S \ref{subsec:outlier}). However, the fraction of these SED ``outliers'' is similar to that in control quasars. This suggests no correlation between the occurrences of candidate optical periodicity and SED anomaly.

\item To explain the abnormally red colours for the seven quasars selected as SED outliers, we have compared the observations with predictions from circumbinary accretion disc models (\S \ref{subsec:sed_location}). 
We find that the SEDs of six out of the seven colour-selected periodic candidates are broadly consistent with theoretical predictions from circumbinary accretion disc models with cutoffs due to a central cavity, where the inner region of the disc is almost emptied by the secondary BH. On the other hand, the minidisc senario, with substantial accretion onto one or both BHs with their own minidiscs, is disfavored, although the limited SED data cannot rule out this possibility entirely given model uncertainties. 

\item We have also considered an alternative scenario of reddening by dust (\S \ref{subsec:dust}). Following the FM formalism \citep[][see also \citealt{Zafar2011,Zafar2015}]{Fitzpatrick1986}, we have modeled the observed SEDs assuming an SMC type extinction curve. The best-fit dust reddening models fit the observations well, with estimated $A_V$ values ranging from $\sim$0.1 to 1.1 mag, which are reasonable for optical quasars.

\item We have considered further evidence for dust reddening based on their WISE colours using the [4.6]$-$[12] vs. [3.4]$-$[4.6] colour-colour diagram (\S \ref{subsec:wise}). Candidate periodic quasars show similar WISE colours to those of control quasars, whereas the subset with abnormally red SEDs is systematically skewed towards the obscured AGN population compared to control quasars, consistent with expectation from dust reddening. Alternatively, a central cavity in the circumbinary accretion disc opened by a secondary BH could also explain the WISE colours.

\item We have discussed possible false positives in the current sample of candidate periodic quasars identified from few-cycle observations (\S \ref{subsec:false_positive}). In particular, we have re-assessed the robustness of the seven candidate periodic quasars with red SEDs by calculating the generalized Lomb-Scargle periodogram based on the public CRTS or PTF light curves. We have carefully examined the significance of any periodogram peak. We have run a large set of simulated light curves that are tailored to the observed variability properties of each quasar. While finding consistent values with the reported periods, none of them exceeds 3$\sigma$ significance, suggesting that most current candidate periodic quasars from few-cycle light curves may be false positives \citep[see also][]{Vaughan2016}.

\item Finally, we have discussed sampling bias driven by optical quasar variability selection (\S \ref{subsec:bias}). Based on the ensemble structure functions (Figure \ref{fig:EDD}), we find that candidate periodic quasars are systematically more variable than control quasars over all timescales. The higher level of variability is largely driven by a selection bias in candidate periodic quasars. Candidate periodic quasars show systematically lower Eddington ratios than control quasars (Figure \ref{fig:EDD}), verifying the known anti-correlation between Eddington ratio and optical quasar variability.

\end{enumerate}

Future work should look for other SED signatures predicted by circumbinary accretion disc models such as hard X-ray excess from stream-disc collisions \citep[e.g.,][]{Roedig2014,Farris2015,Farris2015a,Foord2017,Krolik2019}. While the sample of candidate periodic quasars does not have enough archival X-ray data to test this, one should look for them in the much larger sample of ordinary quasars with archival X-ray observations \citep[e.g.,][]{Civano2012,Coffey2019}. Future work should also search for possible SED outliers to select BSBH candidates independent from the optical periodicity selection, which is still largely subject to false positives given limitations in the current light curve data. Finally, our work motivates the identification of more robust samples of candidate periodic quasars both by significantly extending the baseline coverage of existing samples with continued monitoring and by more carefully assessing the statistical significance of any candidate periodicity.

\section*{Acknowledgement}
%We thank our referee for his prompt and constructive report that helped significantly improve the paper.
We thank M. Sun for help with setting up the binary model, Y. Shen and K. G{\"u}ltekin for helpful discussions, Z. Haiman for useful comments, and an anonymous referee for a quick and constructive report that significantly improved this work.

%%%%%%%%%%%%%%%%%%%%%%%%%%%%%%%%%%%%%%%%%%%%%%%%%%

%%%%%%%%%%%%%%%%%%%% REFERENCES %%%%%%%%%%%%%%%%%%

% The best way to enter references is to use BibTeX:

\bibliographystyle{mnras}

%\bibliography{/Users/zeus/Documents/References/binaryrefs} % if your bibtex file is called example.bib
\bibliography{binaryrefs}

%%%%%%%%%%%%%%%%% APPENDICES %%%%%%%%%%%%%%%%%%%%%

\appendix

\section{Remarks on Individual Candidate Periodic Quasars with Red SEDs}\label{sec:appendix}

\begin{figure}
 \centering
  \vspace*{0cm}
  \hspace*{-0.5cm}
  \includegraphics[width=100mm]{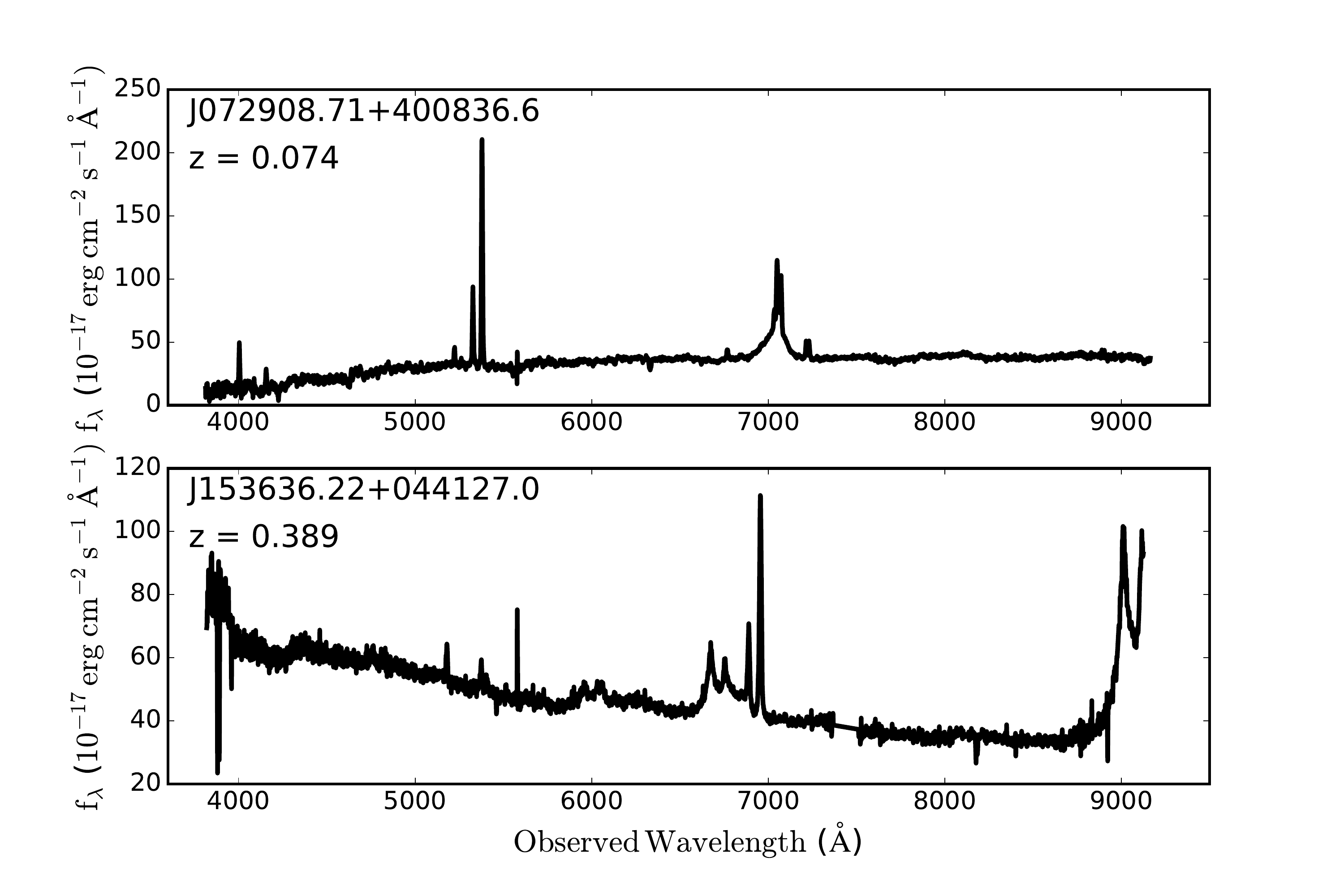}
    \caption{SDSS spectra of two interesting cases in candidate periodic quasars with red SEDs. Upper panel shows a type 1.9 quasar with a strong broad-line component in \ha\ but not in \hb. Lower panel shows a quasar with a double-peaked broad emission line, which has been proposed as a candidate sub-pc SBHB by \citet{boroson09} and suggested as an unusual accretion disc emitter by \citet{chornock10}.  
    }
  \label{fig:spec}
\end{figure}

SDSS J072908.71+400836.6 is a type 1.9 quasar. It has a broad-line component in \ha\ but not in \hb\ (Figure \ref{fig:spec}, upper panel). Its SED and spectrum can be well fit (Figure \ref{fig:cand}) by a composite quasar SED reddened by an extinction curve model (Equation \ref{eq:fm}) with $A_V{\sim}$1.1 mag (Table \ref{table:table2}). Similarly, Mrk 231 also shows a strong broad \ha\ but a weak broad \hb\ and a red continuum. In addition, Mrk 231 is a broad absorption line quasar with UV variability \citep{Yang2018}. \citet{Yan2015} has proposed Mrk 231 as a candidate BSBH whose red SED is driven by a notch in circumbinary accretion disc, while \citet{Leighly2016} suggests that it is a reddened AGN.

SDSS J153636.22+044127.0 is a quasar with double-peaked broad emission lines (with velocity a separation of 3,500 \kms ; Figure \ref{fig:spec}, lower panel). Based on its double-peaked broad emission lines, it was identified by \citet{boroson09} as a candidate sub-pc BSBH system having masses of $10^{7.3}$ and $10^{8.9}$ $M_{\odot}$ separated by $\sim$0.1 pc with an orbital period of $\sim$ 100 years. \citet{chornock10} has suggested that it is instead an unusual double-peaked disc emitter, whose broad-line velocity splitting is driven by rotation and relativistic effects of accretion discs around single black holes \citep[e.g.,][]{eracleous03,lewis10}. \citet{shen10} has suggested that for large broad-line velocity splittings, the two BHs in a BSBH will be too close for their broad-line regions to be distinct, resulting in complex line profiles rather than a clear splitting of the peaks, which does not correspond to binary orbital motion.  G15 has also identified SDSS J153636.22+044127.0 as a candidate BSBH but based on its candidate optical light curve periodicity with a period of $\sim$3 years (in the observed frame), resulting in a $\sim$10 times smaller estimate for the binary separation ($\sim$0.01 pc). Our independent analysis of the periodicity significance level is only $\sim$2$\sigma$ (\S \ref{subsec:false_positive} and Table \ref{table:table1}). The CRTS light curve covers only $\sim$3.5 cycles of the candidate periodicity, which may be too short to reject false positives from stochastic, red noise quasar variability \citep{Vaughan2016}.

%%%%%%%%%%%%%%%%%%%%%%%%%%%%%%%%%%%%%%%%%%%%%%%%%%

% Don't change these lines
\bsp	% typesetting comment
\label{lastpage}
\end{document}